\documentclass[letterpaper]{article}

\usepackage[english]{babel}
\usepackage[utf8]{inputenc}
\usepackage{amsmath}
\usepackage{graphicx}
\usepackage{multirow}
\usepackage[top = 1in, bottom=1in, left=1in, right=1in]{geometry}
\usepackage{multicol}
\usepackage[hang]{footmisc}
\usepackage{fancyhdr}
\usepackage{alltt}
\usepackage{natbib}
\usepackage{url}
\usepackage{color}
\usepackage{pdfpages}

\newcommand{\fn}{\footnote}
\newcommand{\todo}[1]{}
\newcommand{\query}[1]{}
\newcommand{\bi}{\begin{itemize}\itemsep2pt \parskip0pt}
\newcommand{\ei}{\end{itemize}}
\newcommand{\ylicontact}{\url{jbernd@yli-corpus.org}}

\title{The YLI-MED Corpus: Characteristics, Procedures, and Plans}

\author{Julia Bernd \and Damian Borth \and Benjamin Elizalde \and Gerald Friedland \and Heather Gallagher \and Luke Gottlieb \and Adam Janin \and Sara Karabashlieva \and Jocelyn Takahashi \and Jennifer Won}
\date{}

\begin{document}

\includepdf{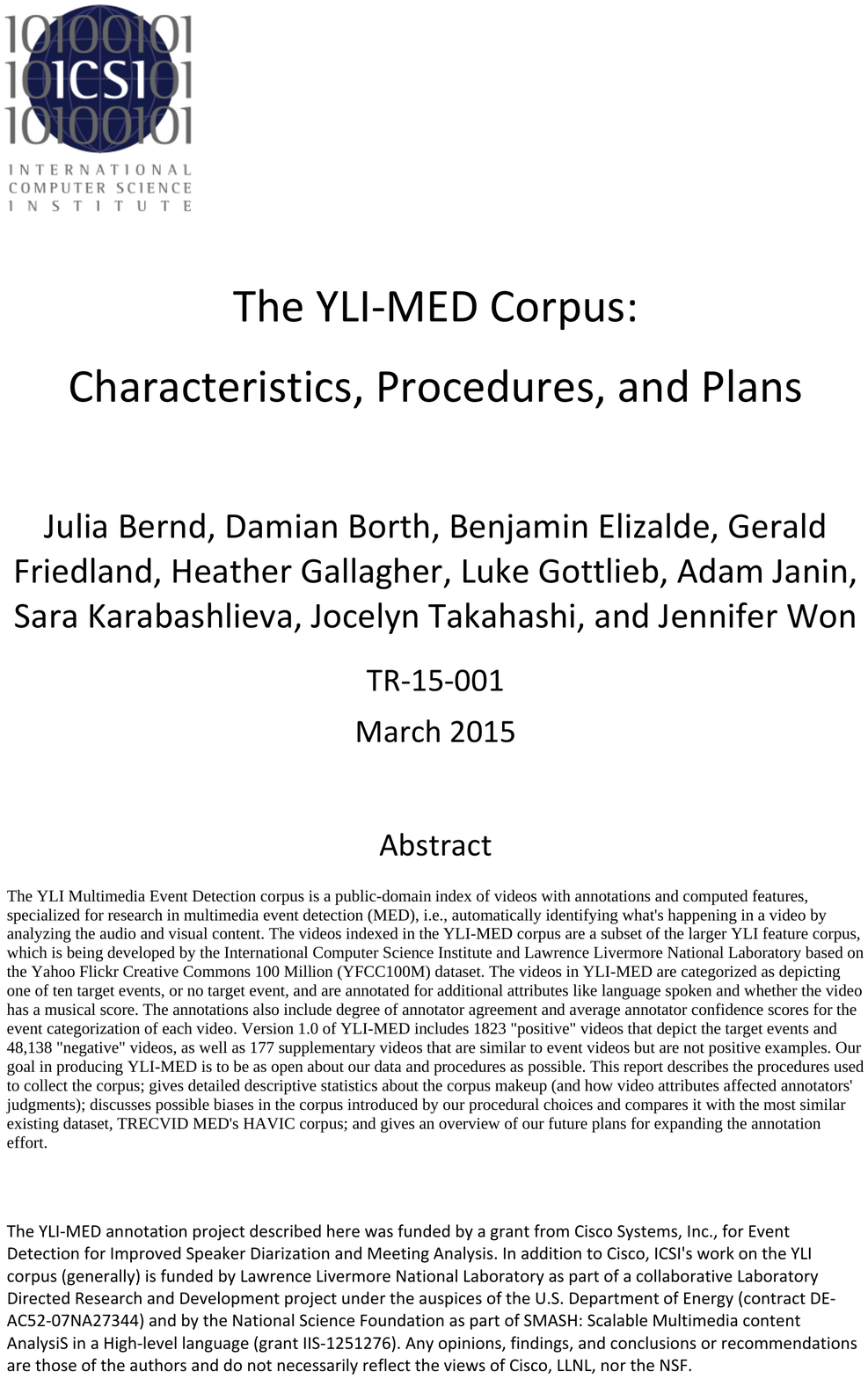}

\newpage
\setcounter{page}{1}

\pagestyle{fancyplain}
\cfoot[The YLI-MED Corpus // ICSI TR-15-001]{The YLI-MED Corpus // ICSI TR-15-001}
\rfoot[\thepage]{\thepage}
\lhead[]{}
\setlength{\headheight}{14pt}

\maketitle

\setlength{\parskip}{1pt}

\tableofcontents

\setlength{\parskip}{2pt}

\section{Introduction}
\label{sec:intro}

YLI-MED is \textbf{the YLI Multimedia Event Detection corpus}, a public-domain index of user-generated videos with annotations and computed features. YLI-MED is specialized for use in the growing field of multimedia event detection (MED) research, i.e., for use in training and testing systems that automatically identify what human-defined events are depicted in a video by analyzing the audio and visual content. It may be obtained at \textbf{\url{http://www.yli-corpus.org/the-yli-med-corpus}}.

The videos indexed in the YLI-MED corpus are a subset of those in the larger YLI corpus,\fn{\url{http://www.yli-corpus.org}} which is being developed by Lawrence Livermore National Laboratory (LLNL) and the International Computer Science Institute (ICSI). YLI is a set of computed audio and visual features and additional annotations (including MED) for the Yahoo Flickr Creative Commons 100 Million (YFCC100M) dataset (\citealt{YFCC,YFCC2014}).\fn{The YFCC100M is available via the Yahoo! Webscope portal at \url{http://webscope.sandbox.yahoo.com/catalog.php?datatype=i\&did=67}. Yahoo makes this data available for free to computer-science researchers, subject to the terms of the Webscope agreement on data use and the Creative Commons licenses chosen by the uploaders.} YFCC100M contains the metadata for 99.7 million photos and nearly 0.8 million videos that have been uploaded to Flickr using a Creative Commons license.\fn{The YFCC100M metadata includes, among other things, user, camera model, timestamp, GPS location information (if available), title, user-supplied tags, text description, machine tags, CC license type, and Flickr media ID. There is also a supplemental file that connects the Flickr media IDs to unique MD5 hashes, which may in turn be used to connect the YLI annotations and features with the metadata and the original media in the YFCC100M.} This unprecendentedly large open-source dataset of user-generated content (UGC) provides unique new opportunities for multimedia research---the more so if it is annotated.

\begin{figure}[t]
\centering
\includegraphics[width=6.5in]{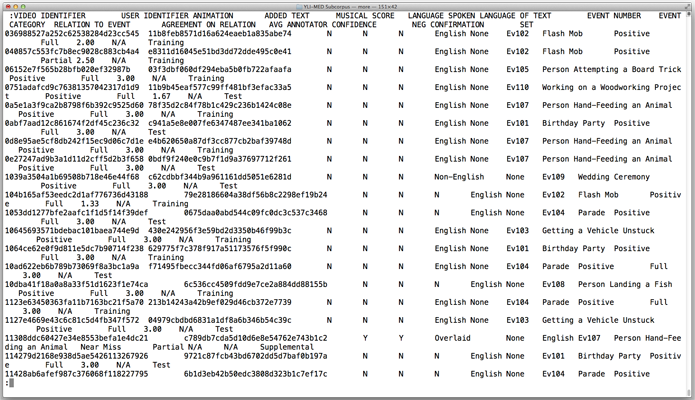}
\caption{A snapshot of the YLI-MED index and annotations.}
\label{fig:screenshot}
\end{figure}

YLI-MED provides such annotations, targeted towards multimedia event detection (see Figure~\ref{fig:screenshot}). The videos in YLI-MED are categorized as depicting one of 10 (so far) target events, or no target event, and annotated for additional attributes of interest, such as language spoken and whether the video has a musical score. The annotations also include degree of annotator agreement and average annotator confidence scores for the event categorization of each video. The index and annotations were compiled at ICSI and are free and open to all MED researchers (or anyone else, for that matter), under Creative Commons License 0.\fn{\url{http://creativecommons.org/publicdomain/zero/1.0/}}

The YLI-MED index includes annotations for 2000 videos split between the 10 target events and ~48,700 videos that do not depict those events, divided into standard training and test sets for researchers wishing to compare results. The events included are: {\it Birthday Party}, {\it Flash Mob}, {\it Getting a Vehicle Unstuck}, {\it Parade}, {\it Person Attempting a Board Trick}, {\it Person Grooming an Animal}, {\it Person Hand-Feeding an Animal}, {\it Person Landing a Fish}, {\it Wedding Ceremony}, and {\it Working on a Woodworking Project}. 

Technical details about the current version of the corpus, including column labels and values for the index file, may be found in the README for the current release. This report provides details of the procedures used to develop the index for Version 1 of the corpus (in \S\ref{sec:procedures}); extensive descriptive statistics about its contents (in \S\ref{sec:character});\fn{Unless otherwise noted, the descriptive statistics refer specifically to Version 1.0, released in September 2014.} a discussion of how YLI-MED compares to other corpora (in \S\ref{sec:comparability}); and a few words about our future plans for YLI-MED (in \S\ref{sec:future}).  

We aim to be as open as possible about the composition of the YLI-MED corpus and the corpus-development process. If you have any questions that are not addressed in this report, please feel free to contact us, via \ylicontact.

\section{Corpus-Collection Procedures}
\label{sec:procedures}

The collection and annotation of the YLI-MED corpus index was accomplished in a series of stages throughout Summer 2014; each of these stages is described in this section. \S\ref{sec:definitions} describes how we chose the target events and researched and developed ``definitions'' for those events. \S\ref{sec:building} describes how the corpus-builders identified and verified videos within the YFCC100M that depicted the target events, and how we annotated them for features of interest to multimedia researchers. In \S\ref{sec:cleanup}, we describe how we calculated annotator agreement and confidence and finalized the index of annotated videos (training, test, and supplemental) to be released (including the ``Alternative'' training set). In \S\ref{sec:negatives}, we describe how we (automatically) selected and (manually) honed a set of random ``negative'' videos that do not depict any of the target events.\fn{The term {\it negative video} is used differently by different corpus providers and users; we mean videos that are not related to the event at all.}

\subsection{Choosing and Defining the Events}
\label{sec:definitions}

For Version 1 of the YLI-MED corpus, we chose to use events that were covered in the TRECVID MED 2011 evaluation (\citealt{TV2011,TRECVID}), organized by the National Institute of Standards and Technology (NIST). The fifteen events used in TRECVID MED 2011 (and also in later evaluations) are listed in Table~\ref{tab:trecvidevents}; the ten asterisked events are also part of YLI-MED Version 1.\fn{The ten were chosen on the basis of coverage in YFCC100M; see \S\ref{sec:collection}.}

\begin{table*}[htp]
\centering
{\renewcommand{\arraystretch}{1.2}
\begin{tabular}{ | l l |}
\hline
Attempting a board trick* & Making a sandwich \\
Birthday party* & Parade* \\
Changing a vehicle tire & Parkour \\
Feeding an animal* & Repairing an appliance \\
Flash mob gathering* & Wedding ceremony* \\
Getting a vehicle unstuck* & Working on a sewing project \\
Grooming an animal* & Working on a woodworking project* \\
Landing a fish* & \\
\hline
\end{tabular}}
\caption{The fifteen events for the TRECVID MED 2011 Evaluation (\citealt{TV2011}).}
\label{tab:trecvidevents}
\end{table*}

One of the main goals of MED research is to produce methods that are general enough to be effective across different datasets. TRECVID MED has been a dominant organizing factor in this field for several years, and much of the research on MED has been connected with it. Having a video corpus that parallels that used by TRECVID in 2011, categorized for the same target events but collected differently and including, at least in part, different videos, provides TRECVID researchers with a specific basis for testing the generalizability of their systems and methods. For researchers new to the field who do not have access to the TRECVID data,\fn{Access to the TRECVID MED data provided by the Linguistic Data Consortium is tied to participation in the evaluation, and even for participants, use of the data for other purposes is restricted.} the open-source YLI-MED provides a means to compare their work to the state of the art. In addition, training classifiers or detectors on different datasets for the same event provides researchers with a new dimension for experimentation, probing the factors that lead to the range of results a given system can produce. 

As with our choice of events, we largely based our general definition of an {\it event} on NIST's published definition (\citealt{TV2011web}). However, the YLI-MED corpus definition of an {\it event} differs from NIST's in one important respect. While NIST and the Linguistic Data Consortium (LDC), which developed the source dataset HAVIC (\citealt{HAVIC}), define an {\it event} (for their purposes) to always have a human protagonist, we explicitly specified that the protagonists could be animals, robots, ghosts, etc.\ as well as humans. However, the nature of the target events (which all centered around intentional human-type activities) meant that those animals, etc.\ would still have to be acting as agentive ``people'', i.e., anthropomorphized in some way.\fn{See \S\ref{sec:comparability} for for further discussion of the differences between the procedures used to collect HAVIC and YLI-MED, biases these differences may have introduced in each corpus, and their comparability as a basis for developing automated event-detection systems.}

We thus adopted the following definition of {\it event}, which is quoted with modifications from NIST's definition in \citealt{TV2011web}:

\noindent \begin{quotation} An event “is a complex activity occurring at a specific place and time; ... consists of a number of ... actions, processes, and activities that are loosely or tightly organized and that have significant temporal and semantic relationships to the overarching activity; and is directly observable [via at least one sensory modality].” \end{quotation}

Our goal in developing the corpus was to include videos from a broad cultural range, i.e., videos that represented what a wide selection of people across demographic groups and geographies would consider to constitute that event (and consequently would consider a valid retrieval result for a search on that event, in a classifier-based application).\fn{The breadth of factors that go into deciding what counts as the ``same'' event across cultures varies, of course. For example, while the distribution of animal life varies quite a bit, the process of a human feeding any animal is recognizable from fairly similar evidence across cultures. On the other hand, a ceremony is by its nature a culturally specific procedure, so a wedding ceremony may take many different forms in different cultures---however, most cultures do have some type of prescribed ritual that results in two or more of the participants being married according to some (variable but recognizable) social, religious, and/or legal standard. In contrast, a flash mob is generally viewed as a more culturally specific phenomenon, so it is more difficult to say whether similar activities across cultures count as the ``same'' type of event.} The corpus-builders (the collectors and verifiers) were therefore told to ``cast a wide net'' and given instructions and training in how to best achieve our twin goals of cultural variety and conceptual unity.

In particular, to ensure a result that was both consistent and broad, staff researchers at ICSI developed event ``definitions'' and decision rubrics to guide the corpus-builders in deciding whether a given video counted as depicting the target event, based on research into how Internet users think about those event categories. These definitions are included here in Appendix~\ref{sec:defsapp}. While it may seem counterintuitive, we viewed giving specific, evidence-based definitions and rubrics as the best way to achieve breadth; in other words, as the best way to prevent corpus-builders from applying too narrow a standard to decide what counted as a valid example, based on their own definition of the terms.\fn{If we had used an approach that involved a large number of collectors and verifiers from a wide range of backgrounds working on each event, the definitions might have been less necessary. However, our workflow and resources dictated that the work be divided up among a small number of people (the majority of the videos were collected by one person), requiring a more proactive approach to avoiding collector bias.} However, the definitions and instructions did impose some limitations on the corpus-builders' interpretation of the event names. In order for the YLI-MED corpus to be used to produce general models of each event that will fit the expectations of a regular Internet user, the choice of videos required some specific rubrics for deciding what belongs in a category, among all of the possible results returned from a given tag search.

In sum, our overall approach in developing the event definitions aimed to be descriptive (``What do people think a {\it birthday party} is?'') rather than prescriptive (``How should people use the phrase {\it birthday party}?''). We gathered evidence about what Internet users might consider to belong to the target event categories---and what they might expect or want to find if they looked for videos using those terms---and abstracted over it to develop the definitions. The evidence came from a number of different types of sources, including:

\bi
\item Curated sources of information with traditional authority, such as how-to's and histories. These provided information about where expert practitioners explicitly drew the category boundaries and why (and what the range of variation was among such sources), and what types of images the experts considered most illustrative.\fn{We do not by any means assume that most Internet users will necessarily have the same category structure as expert practitioners; the expert sources simply provide one possible perspective (which is, of course, a potential source for other Internet users' conceptualizations of events, as users' searches may---or may not---be based on perceptions about expert terminology).}
In particular, how-to's provided information about the temporal scope of an event (for example, whether jacking up the car should be included as part of the {\it Changing a Tire} event).
\item Lexical resources, both prescriptive (such as Merriam Webster\fn{\url{http://www.merriam-webster.com}}) and descriptive (such as FrameNet\fn{\url{http://framenet.icsi.berkeley.edu}}). These clarified the range of meanings of the words in the event name (not all of which would be relevant) and provided insight into how the specific words in the event name might contribute in evoking the semantic frame(s) relevant to that event (\citealt{fillmore1985}).\fn{One of the challenges in ``defining'' an event is to distinguish between defining the specific English words or phrases used in the event name and describing the target event named by those words. We started with NIST's event names, which were for the most part well-formulated to be as precise as possible, but the events referred to could also have been named---and thus could be searched for---using other words even in English, not to mention other languages, which may not have exact translations for the English words being used. For example, there is substantial overlap between the events that could be described using the English word {\it parade} or the Spanish words {\it parada} or {\it desfile}, but there are also words that might be covered by one of those three terms but not the other. Our task was to define the event {\it Parade} in such a way as to cover a conceptual space that would be recognizable as coherent to speakers of either English or Spanish.}
\item Semi-curated sources with open contributorship, such as Wikipedia and Wikihow. These provided a view of what a broader sector of Internet users considered to constitute various categories (as compared to curated expert sources), while still focusing on what people think counts as an example of the category when they are explicitly thinking about it. In addition, these sources frequently highlighted the explicit and implicit range and conflicts in what different groups of people considered to be part of a given category. (For example, the Wikipedia entry on {\it Flash mob}\fn{\url{http://en.wikipedia.org/wiki/Flash_mob}} has some explicit discussion of conflicts over what counts as a ``real'' flash mob, and there is a lengthy debate in the Talk page for the entry.) 
\item Example image and video searches using the words or phrases in the event name and related words or phrases. Identifying the scenes that came up most commonly showed what the posters of media most frequently use those tags for\fn{Reinforced, perhaps, by tag-suggestion systems based on exactly such data.} and at the same time provided a quick check of what existing search-engine analysis had identified as most likely to be relevant. Searching on a number of related words and phrases and comparing results provided insight into how important (or not) the specific words in the event name were to circumscribing the nature of that event.
\ei

In addition to the aforementioned goals, we also wanted to maintain comparability with the TRECVID MED datasets for the same events. Our definitions therefore tried to account for what NIST and the LDC considered to constitute those event categories, at least at a basic level.\fn{In order to avoid using proprietary elements of the LDC's work, preparation of the definitions and instructions for YLI-MED was led by a researcher who had not participated in TRECVID and had not seen the LDC's unpublished definitions.} The definitions and rubrics were, however, formulated differently from the LDC's in a number of ways. For example, we assumed it would be relatively easy for corpus-builders to make decisions about the most prototypical instances of a given event type, and more difficult to make decisions about the less prototypical instances (at the periphery of the category).\fn{Our overall approach to collecting a corpus of videos that fit into a particular category (i.e., that depict a specified type of event) assumes non-classical category theory (\citealt{rosch1977}, \citealt{mr1981}, \citealt{lakoff1987}), i.e., assumes that people think of categories in terms of a core or central prototype for what they expect a member of that category to be like, with less prototypical members being viewed as more peripheral to the category, and that category boundaries and belongingness are thus gradient. (As opposed to classical category theory, wherein categories are assumed to have strict boundaries defined by the necessary and sufficient conditions for membership.) To the degree possible, rubrics were framed explicitly in those terms, rather than in terms of hard-and-fast conditions.

To make things more complicated, in general, humans evaluate prototypicality along multiple dimensions at once. While instances that deviate from the central prototype along some dimension or another might still be considered quite fine members of their category by most people, as the dimensions of nonprototypicality add up, it becomes less and less likely to be considered a good example of the category---but this ``additivity'' is not linear nor (as yet) predictable from first principles, so it presents a very interesting challenge for building systems to emulate humans' category judgments.} The bulk of the decision rubrics were therefore devoted to what to do with the boundary cases---for example, for {\it Person Hand-Feeding an Animal}, ``Any type of solid or liquid food can be provided (including `treats'), whether or not it is actually appropriate for the animal's diet.''---rather than the easy cases that made up the bulk of the videos they found.\fn{Again, this approach was dictated by the particular structure and skills of our corpus-building team, and would not necessarily have been the best approach for a larger or non-in-house group.}

The difficult cases for the corpus-builders might have to do with the fuzzy boundaries of the event category itself (such as {\it parade} or {\it flash mob}) or might have to do with an aspect or product of the event (such as {\it project} or {\it sandwich}). At worst, some of the categories were multi-centered and/or contested. {\it Flash mob} is both; as we noted above, reasonable people frequently disagree about what counts as a ``real'' flash mob, and there are also multiple prototypes. In such cases, the problem was noted explicitly and the corpus-builders were told to allow the definition to stretch broadly (see Appendix~\ref{sec:defsapp}). For the most part, the corpus-builders seem to have stuck to the definitions and rubrics provided (based on our research on other Internet users' view of those events) even where the rubrics deviated from their individual views of what did and did not belong in the category. In most cases, any idiosyncratic interpretations that influenced video choice were balanced out by the fact that each video was annotated three times (see \S\ref{sec:collection}). In the few cases where all of the corpus-builders seemed to be deviating from the rubric provided, the staff researchers took this as potential evidence that the rubrics might need to be narrowed or broadened. 

Generally, the staff researchers and the corpus-builders interacted throughout the process and videos were spot-checked as they were being gathered. As a result, we made a few minor changes to the event definitions mid-process, usually to clarify unclear wording.

{\it A Note on Names:} The event names for the YLI-MED do not necessarily reflect those used for TRECVID MED. In a couple of cases, the TRECVID names were ambiguous, so we used names that reflected how we resolved that ambiguity (for example, {\it Person HAND-Feeding an Animal}, where the TRECVID MED name {\it Feeding an Animal} was not clear as to whether it included, for example, leaving food in an animal's dish or stall when the animal wasn't present). In addition, TRECVID MED event names did not specify anything about the protagonist, as NIST's overall definition of an {\it event} made clear that it would always be human. Since our definition of an {\it event} does not specify any such thing (as we plan to later add events that do not have human protagonists), it was necessary to reformulate some events as {\it Person Doing X} (though again, a ``person'' can be non-human). However, it should be noted that the corpus-builders were initially given the TRECVID MED event names.

The final list of event names and numbers for YLI-MED is given in Table~\ref{tab:finalevents}.

\begin{table*}[htp]
\centering
{\renewcommand{\arraystretch}{1.2}
\begin{tabular}{ | l l | l l |}
\hline
Ev101 & Birthday Party & Ev106 & Person Grooming an Animal \\
Ev102 & Flash Mob & Ev107 & Person Hand-Feeding an Animal \\
Ev103 & Getting a Vehicle Unstuck & Ev108 &Person Landing a Fish \\
Ev104 & Parade & Ev109 & Wedding Ceremony \\
Ev105 & Person Attempting a Board Trick & Ev110 & Working on a Woodworking Project \\
\hline
\end{tabular}}
\caption{The ten events in Version 1 of the YLI-MED corpus.}
\label{tab:finalevents}
\end{table*}

\subsection{Event Corpus Collection, Annotation, and Verification}
\label{sec:building}

\subsubsection{Collection Procedures}
\label{sec:collection}

An index of candidate videos was compiled by searching Flickr for Creative Commons videos depicting the target events, using the searchable tags, descriptions, etc.\ provided by the uploaders.\fn{The corpus-builders (collectors and verifiers) all worked in-house for ICSI. Copies of the original instructions provided to the video collectors and verifiers may be obtained by emailing \ylicontact.} Searches were filtered using the parameters used to select the videos for the full Yahoo Flickr Creative Commons 100 Million dataset.\fn{The parameters were: All Creative Commons videos; SafeSearch On; Content Type includes videos, screencasts, and animation/CGI; and posted before 3/21/2014 (the final date for YFCC100M collection).} The corpus ``collection'' process was thus in effect a process of identifying a subcorpus of the YFCC100M to use for YLI-MED, but it was conducted using the Flickr website in the interests of expediency.

The corpus collectors were asked to find 150-250 videos that depicted each of the target events. In addition to these ``positive'' examples, they were asked to also index videos that were evidential ``near misses'' for depicting the event or were closely ``related'' to the video without actually depicting it (though there was no numerical target for these categories).

The collectors recorded in a spreadsheet %
the URL of the video and information about the event category and the video's relation to that event, i.e., whether they considered the video to be a positive instance, a near miss, or a related video. For positive videos, they also recorded their confidence about whether the video was indeed a positive example of the target event, on a scale from 1 to 3, with 1 being ``Not sure'' and 3 being ``Absolutely sure''. They also recorded information about production and post-production features and languages, as described in \S\ref{sec:annot}.

The collectors (and later the verifiers) were given the following criteria for determining Relation to Event: 

\begin{quotation}

\noindent `Positive Instance' Criteria:\fn{These criteria are analogous to the LDC's ``Sufficient Evidence'' and ``Reasonable Viewer'' rules used by their data scouts collecting videos for HAVIC (\citealt{HAVIC}).}
\bi
\item There is sufficient evidence in the visual and/or audio cues to verify that the target event is occurring;\fn{The corpus-builders were also told that user-supplied textual metadata could be used as evidence only if the audio and visual cues were already very strong, and metadata was only needed to disambiguate between close possibilities.} AND
\item What happens in the video is a close enough match to the definition that you think most reasonable viewers would consider it to be an instance of that event.
\ei 

\noindent Note: The video doesn’t needs to contain the whole prototypical series of steps/subevents within the event from start to finish to be considered an instance, but it needs to contain enough of those steps to be uniquely identifiable.

\smallskip\noindent `Near Miss' Criteria:
\bi
\item The video bears a significant similarity to positive instances of the event, but the preponderance of the evidence suggests that it is actually an instance of some other event; OR 
\item The evidence suggests that the video could be considered an instance of the target event by some reasonable viewers, but it is nonprototypical enough (along some dimension) that many reasonable viewers might disagree that it fits the definition.
\ei 

\noindent `Related' Criteria:
\bi
\item The video is related to the target event category in that it has some prototypical elements that one might expect to see or hear in a video of the target event, but it is definitely not an instance of the target event, nor even close enough to be considered a `Near Miss'.
\item For example, `Related' videos might include people talking about that event (e.g., talking about how to change a tire), preparing for or cleaning up after such an event, using or showing off items that are prototypically associated with that event (e.g., for \textit{Working on a Sewing Project}: using a needle and thread to sew up a wound, or an unboxing video for a new sewing machine), etc.
\ei 

\end{quotation}

Videos were excluded from the candidate index if they were shorter than 3 seconds, had no audio track (or no visual track), or contained possible copyright violations or material that shouldn't have turned up in SafeSearch. Videos could theoretically be listed under multiple event categories, but there are not actually any such cases in YLI-MED v.1.

The video collectors recorded all of the search terms used, with notes as to which search terms were most and least fruitful (or whether they produced any results at all). They were given initial general suggestions about how to come up with good search terms, and more specific suggestions if they were having trouble finding videos for an event. Generally, collectors were encouraged to try search terms in different languages, in order to increase cultural breadth (though still somewhat limited by the inexactness of translation), but in practice, they tended to turn to other languages only when they had indexed all of the videos they could find using English search terms (which happened with only some of the events). However, even where collectors tried search terms in many languages, the vast majority of results came from the English searches.\fn{This to some degree reflects the tendency for Flickr tags to be in English or English and another language (\citealt{folks}). (Though this is less the case for titles and descriptions, which were also included in the searches.) However, it may also beg the question of how effective collectors could be at coming up with appropriate search terms in other languages (the way it would actually be said). While linguists on the staff research team assisted with search-term brainstorming, there were, of course, likely some pertinent gaps. In addition, while Flickr does well at matching inflected search terms with other variations on the same root in English ({\it changes} gets results with {\it change}, {\it changing}, {\it changed}, etc.), its inflection dictionaries for other languages do not seem to be nearly as robust. This limits a searcher's ability to efficiently run through likely possibilities in a language with which they are not familiar---especially for highly inflected languages.} This distribution varied somewhat across events, depending on cultural generality vs.\ specificity (the U.S.\ English term {\it flash mob} is generally borrowed into other languages when speakers want to describe that specific phenomenon) and cultural associations (probably because the U.S.\ is seen as a cultural hub for skating, skateboard videos are often tagged {\it skate} even when the rest of the metadata is in other languages). However, the variation was surprisingly less than we had expected.\fn{A full record of the search terms used to find videos in each category may be obtained by emailing \ylicontact.} 

In five cases, the video collectors were not able to find enough videos to make up reasonably sized training and test sets for classifier development, even with extensive search-term assistance from linguists on the research team, leaving the ten events listed in Table~\ref{tab:finalevents}. \fn{The number of videos found for the five events before the decision was made to exclude them were as follows: {\it Changing a Vehicle Tire} 11; {\it Making a Sandwich} 30; {\it Parkour} 58; {\it Repairing an Appliance} 1;{\it Working on a Sewing Project} 43. In some cases, the corpus-building team tried many less-promising search terms and many languages to achieve that number; in some cases, they tried only the more obvious English terms, but results were few enough from those efforts that it was immediately apparent we would not meet the target.}

\subsubsection{Annotation for Non-Event Characteristics}
\label{sec:annot}

In addition to event category, relation to that event, and confidence in their categorization, the corpus-builders also noted the following information about the content, production, and post-production characteristics of each reviewed video.\fn{The positive instances, near misses, and related videos are referred to collectively in this report as the ``reviewed videos'', to distinguish them from the (ideally) truly unrelated ``negative'' videos, most of which were not individually reviewed. (And even for the subset of the negatives that was checked to confirm their designation as negatives (see \ref{sec:negatives}), they were not fully annotated and reviewed by multiple people.)} Except as noted in \S\ref{sec:verification}, these annotations were made by the person who initially collected each video for the corpus.

\bi
\item Animation: Whether the video includes (or consists of) animated sequences, including CGI.
\item Added text: Whether the video includes text added in post-processing, such as title screens or subtitles. (This does {\it not} refer to incidental text within the scene depicted in the recording, such as street signs or logos; the latter information was not tracked.)
\item Musical score: Whether the video has a music track added in post-processing. Subcategories:
\bi
\item Score only = original audio has been completely replaced;
\item Interspersed = original audio is still present for some segments of the video, but replaced for other segments;
\item Overlaid = original audio is still included, with music track overlaid on top (for some segments or for the whole video).
\ei
\item Language spoken: Name(s) of any language(s) spoken or sung in the audio (including in music tracks). If the corpus builders could not identify a clearly non-English language (in spoken language or added text), they used the generic label `Non-English'. If a spoken language was not even clear enough to be identified as English or non-English, it was labeled `Unintelligible'.
\item Language of text: Any language(s) used for compositional elements added in post-processing, such as titles or subtitles. (Does {\it not} refer to incidental text within the scene depicted in the recording.)
\ei

A note of caution to users: Corpus-builders were instructed that they could use the language of the metadata (only) to support identification of the language spoken in the video, but a spot-check reveals that they may in a few cases have filled in the ``Language Spoken'' column with the metadata language when there was no spoken nor sung language in the video itself. We plan to correct this error in future versions of the corpus.

Our choices of annotations were based on the attributes most immediately relevant to video analysis and event detection; while there are many other features that would also be interesting to annotate (and that we hope to annotate in future), we deemed these the most urgent for researchers investigating variation in classification and detection performance. We considered it particularly necessary to track whether videos were animated because YLI-MED differs from HAVIC (the most comparable dataset) in our choice to include animation and CGI at all. 

\subsubsection{Verification}
\label{sec:verification}

After we had collected (most of) the candidate videos (positives, near misses, and related) for the YLI-MED corpus, we began a process of verification, in which each video was checked by two additional people. To encourage verifiers to actually log their disagreement (where they disagreed), they were asked to record separately their judgments on each video's relation to the event, and their confidence in their judgment for positive examples (rather than simply recording their confirmation/disconfirmation of the collector's annotation---though they could see the collectors' and each other's judgments). They were also asked to verify the collectors' judgments for the additional annotations (post-production and language), but in the latter cases, they were only asked to log their judgments if they differed from or added to the collectors'.

The instructions for the verifiers largely mirrored those for the collectors (with the exception that they could designate a video a ``negative'' example, i.e., not related to the event at all). The corpus verifiers were also asked to add or clarify some annotations, namely, to clarify the type of music track and to annotate the language of added text. The verifiers also added subcategory designations for some of the {\it Person Attempting a Board Trick} videos (skateboard, snowboard, or surfboard) that were used to create the supplementary Ev105 index (described in \S\ref{sec:extratrick}).

\subsection{Event Set Cleanup and Division into Training/Test}
\label{sec:cleanup}

During the verification phase, the index of candidate videos was double-checked for duplicates and checked against the full YFCC100M index, leaving us with an index of 2274 videos that could potentially be included in the event sets. Annotations were also cleaned up and regularized as necessary at this point.

After verification was complete, we determined the consensus relation-to-event, agreement among annotators, and the average annotator confidence for each video, as described in \S\ref{sec:iaa}, pruned the index to reduce bias towards high-uploading users, as described in \S\ref{sec:userbias} and \S\ref{sec:extratrick}, and divided the final list into Training, Test, and Supplemental sets for each event, as described in \S\ref{sec:splitting} and \S\ref{sec:nonpos}.

\subsubsection{Consensus Relations/Removing Unagreeable Videos}
\label{sec:iaa}

To determine consensus on whether each video was a positive instance, a near miss, or a related video, we used agreement scores of 2/3 or 3/3 between the three annotations for Relation to Event (with no preference to the original collector vs.\ the verifiers).\fn{It should be noted that references to ``annotator agreement'' throughout this report refer to these per-video scores, not to overall inter-annotator reliability ratings. The contingency of the verifiers' judgment on that of the collectors' (i.e., verifiers were only presented with videos the collectors had already judged as examples of, or at least related to, the event) make it difficult to appropriately calculate overall reliability.} We recorded cases where 2/3 (two out of three) annotators agreed on the relation as ``Partial Agreement'' and cases where all 3 agreed as ``Full Agreement''. Table~\ref{tab:relbyagr-init} shows levels of agreement for all of the videos collected, broken down by the consensus relation to the event.\fn{Note that all of the statistics in this section were calculated on the initial index of potential videos to be included, before it was pruned to reduce user bias, so numbers do not match the corpus as released.} Nearly all of the videos on which all of the annotators agreed---99\%---were those for which the consensus was that they were positive instances, while many of the videos on which they disagreed were considered near misses, related videos, or even negatives by two out of the three annotators.\fn{There are no cases of negative videos with full agreement, as the original video-collectors only indexed videos they considered to be positives, near misses, or related, while negatives were simply ignored.}

\begin{table*}[htp]
\centering
{\renewcommand{\arraystretch}{1.2}
\label{tab:relbyagr-init}
\begin{tabular}{ | p{1.4375in} | p{1.4375in} | p{1.4375in} | p{1.4375in} | }
\hline
Agreement Level & Agreement as \% of Initial Index & Consensus Relation to Event & Relation as \% of Agreement Level \\ \hline\hline
No Agreement & 2.1\% & \multicolumn{2}{|l|}{N/A (all three different)} \\ \hline\hline
\multirow{4}{*}{Partial Agreement} & \multirow{4}{*}{13.9\%} & Positive & 43.4\% \\ \cline{3-4}
 & & Near Miss & 30.1\% \\ \cline{3-4}
 & & Related & 18.0\% \\ \cline{3-4}
 & & Negative & 8.5\% \\ \hline\hline
\multirow{3}{*}{Full Agreement} & \multirow{3}{*}{84.0\%} & Positive & 98.7\% \\ \cline{3-4}
& & Near Miss & 1.0\% \\ \cline{3-4}
 & & Related & 0.3\% \\ \hline
\end{tabular}}
\caption{Percentage of potential videos for the corpus achieving each level of agreement. (See Table~\ref{tab:relbyagr-final} for parallel statistics for the released corpus.)}
\end{table*}

In the vast majority (85\%) of cases where two out of three (2/3) annotators judged a video to be a positive instance, the remaining annotator judged it to be a near miss. Conversely, for the majority of videos (79\%) judged as near misses by 2/3 annotators, the remaining annotator judged it to be positive. For those judged by 2/3 annotators to be related videos, a slight majority (54\%) were judged as near misses by the remaining annotator, but the effect was not as strong as for confusion of positives and near misses. Interestingly, among videos judged to be negative (totally unrelated) by 2/3 annotators, a majority (70\%) had been judged positive by the third annotator (necessarily the initial collector); however, since there were many fewer videos in this category and the patterns were more constrained, it is not clear how much can be made of this. There were no significant patterns in judgments on the videos for which there was no agreement at all; they were split approximately evenly between the four possible combinations. %

Table~\ref{tab:agrbyrel-init} shows the proportion of videos with full vs.\ partial agreement for each type of relation to the target event, for all of the potential corpus videos for which there was a consensus on the relation.\fn{Again, since the initial collectors discarded negatives, there are no full-agreement negatives; in addition, once they are decreed negative, they no longer have a relation to a specific event. Since the distinction is therefore not relevant, the (few) partial-consensus negatives are not included in Table~\ref{tab:agrbyrel-init} nor Table~\ref{tab:agrbyevent-init}.} 

\begin{table*}[htp]
\centering
{\renewcommand{\arraystretch}{1.2}
\begin{tabular}{ | p{1.4375in} | p{1.4375in} | p{1.4375in} | p{1.4375in} | }
\hline
Agreement Level & Positive Instances & Near Misses & Related Videos \\ \hline
Partial Agreement & 6.8\% & 82.6\% & 91.9\% \\ \hline
Full Agreement & 93.2\% & 17.4\% & 8.1\% \\ \hline
\end{tabular}}
\caption{Level of agreement for each type of relation to event category among candidate videos collected for the corpus. (See Table~\ref{tab:agrbyrel-final} for parallel statistics for the released corpus.)} 
\label{tab:agrbyrel-init}
\end{table*}

Table~\ref{tab:agrbyevent-init} breaks down the agreement levels by event category. It is worth noting that there is very significant variation in agreement for the different events (p $<$ 0.0001, chi-square), from 61\% full agreement for Ev102 {\it Flash Mob} to 95\% full agreement for Ev104 {\it Parade}---interestingly, two events that can have rather similar audio and visual profiles. This variation largely either parallels prior observational evidence about which of the events involve contested categories (according to the degree of explicit argument about what ``counts'' as a category member in online information sources) or reflects issues explicitly raised by annotators regarding evidence for judgment (or both). The latter often related to how much inference has to be done about the protagonists' intent (i.e., how likely it might be that they had other reasons for performing the actions we can directly observe). (Potential sources of variation are discussed further in \S\ref{sec:charovw}.) 

\begin{table*}[htp]
\centering
{\renewcommand{\arraystretch}{1.2}
\begin{tabular}{ | p{0.5in} | p{2.25in} | p{0.85in} | p{0.85in} | p{0.85in} | }
\hline
Event \# & Event Name & No Agreement & Partial Agreement & Full Agreement \\ \hline
Ev101 & Birthday Party & 0.4\% & 6.4\% & 93.2\% \\ \hline
Ev102 & Flash Mob & 5.4\% & 33.7\% & 61.0\% \\ \hline
Ev103 & Getting a Vehicle Unstuck & 3.2\% & 19.8\% & 77.0\% \\ \hline
Ev104 & Parade & 0.8\% & 4.2\% & 95.1\% \\ \hline
Ev105 & Person Attempting a Board Trick & 1.5\% & 9.1\% & 89.4\% \\ \hline
Ev106 & Person Grooming an Animal & 0.7\% & 17.2\% & 82.1\% \\ \hline
Ev107 & Person Hand-Feeding an Animal & 3.3\% & 6.2\% & 90.5\% \\ \hline
Ev108 & Person Landing a Fish & 3.9\% & 22.9\% & 73.2\% \\ \hline
Ev109 & Wedding Ceremony & 1.3\% & 5.6\% & 93.2\% \\ \hline
Ev110 & Working on a Woodworking Project & 2.2\% & 16.3\% & 81.5\% \\ \hline
\end{tabular}}
\caption{Level of agreement per event among all candidate videos collected for the corpus (positive, near miss, and related). (See Table~\ref{tab:agrbyevent-final} for parallel statistics for the released corpus.)} 
\label{tab:agrbyevent-init}
\end{table*}

Once the agreement scores had been assigned, we removed from the index any videos for which the consensus relation was Negative (i.e., where two out of three annotators said it was not related to the event at all); this comprised 1.2\% of the initial videos gathered. We also removed any videos for which there was no agreement at all on how the video related to the event category; this comprised 2.1\% of the initial videos gathered. This resulted in a total of 2199 videos in the potential video index, of which 2022 (92.0\%) were positive examples of the events, 115 (5.8\%) were near misses, and 62 (2.8\%) were related videos.

We also calculated average annotator confidence scores for each of the positive-instance videos. Where only two of the annotators had judged the video a positive instance (partial agreement), we calculated the average annotator confidence only between the two who had judged  it positive in the first place.\fn{For additional statistics on characteristics of the initial index of candidate videos, contact \ylicontact.}

\subsubsection{Reducing User Bias}
\label{sec:userbias}

In the early part of the corpus-collection process, we did not limit the number of videos that could be gathered from a single user, as we did not yet know how many videos we would be able to find for each event at all.\fn{By {\it user} or {\it uploader}, we mean a Flickr account; of course, some individual users may have multiple accounts.}   However, once we had compiled an early version of the index, we examined the distribution of videos contributed by individual users, and found that there were several events that were skewed towards particular users, including three events for which one user had supplied more than 10\% of the videos.

We recognized that corpus users might have concerns about classifiers potentially overfitting to particular users supplying the training data, or about evaluation scores being skewed by lack of variety in the test data. To manage the risk of user bias, creators of datasets from online platforms like Flickr and YouTube often set a maximum number of images or videos per user (e.g., \citealt{ulges2012linking, borth2013large}); we elected to follow suit. There is some published data to support this common practice, for example the use of user clustering for Social Event Detection (overview in \citealt{SED}), where user biases are actively leveraged, and a preliminary exploration of visual concept detection using data from multiple videos in a uniformly produced series (\citealt{Adrian}). However, hard experimental data on the sources and consequences of user bias is limited; see \S\ref{sec:extratrick} for an invitation (with dataset) to future research.

As we were completing the initial index of candidate videos, we therefore collected additional videos for some of the events. Then, after the cleanup steps described in \S\ref{sec:iaa}, we pruned out an additional 199 positive-instance videos from the high-uploading Flickr users, leaving us with 1823 positive videos for the final corpus to be released.\fn{This number refers to YLI-MED Version 1.0. Versions 1.1 and later have slightly fewer videos, as videos that were removed from Flickr have been subtracted from the index. Unless otherwise noted, all counts and statistics in this report apply to Version 1.0.}

For most of the event categories, we removed random videos from the high-uploading users until there were no more than five videos per user. The goal was that neither the training set nor the test set should have more than 5\% of its videos being from a single user. (For events with fewer than 200 videos, this required manipulation of which users went into the training vs.\ the test sets; see \S\ref{sec:splitting}.) For some of the smaller event categories, where removing so many videos would leave a test set of under 50 videos, we allowed the training sets to have up to 10\% of their videos being from a single user. (All test sets had $\leq$ 5\% from any given user.) For most events, this required removing only a few videos; the major exception was {\it Person Attempting a Board Trick}, about which more is said in \S\ref{sec:extratrick}.  

Table~\ref{tab:usersplit-comp} shows how many videos were removed for each event. It also shows the user distribution for the whole event set before and after the pruning step, in terms of the maximum contribution from any user, the percentage of videos coming from the ten highest contributors collectively, and the (minimum) percentage of users contributing half of the videos in the event set.\fn{More extensive user distribution statistics for the final corpus are in \S\ref{sec:peruser}.} (The final distributions of for the training and test sets separately are given below in Table~\ref{tab:usersplit}.) It is worth noting that, even after this process, the patterns of distribution were quite different for the different events. At the high end, Ev108 {\it Person Landing a Fish} was heavily skewed towards a few prolific contributors, with 38\% of the videos being contributed by just 10 users (12\% of the total contributors), while many of the other events were more evenly distributed, with the top 10 users contributing around 20\% to 25\% of the videos.

\begin{table*}[htp]
\centering
{\renewcommand{\arraystretch}{1.2}
\begin{tabular}{ | p{0.4in} | p{1.25in} || p{0.35in} | p{0.35in} || p{0.4in} | p{0.4in} || p{0.4in} | p{0.4in} || p{0.4in} | p{0.4in} | }
\hline
Event & \multirow{3}{*}{Event Name} & \multicolumn{2}{|c||}{Number} & \multicolumn{2}{c||}{Maximum \%} & \multicolumn{2}{|c||}{\% of Vids from} & \multicolumn{2}{|c|}{\% of Users per} \\ 
\# & & \multicolumn{2}{|c||}{of Videos} & \multicolumn{2}{c||}{per User} & \multicolumn{2}{|c||}{Top 10 Users} & \multicolumn{2}{|c|}{50\% of Videos} \\ \cline{3-10}
 & & Pre & Post & Pre & Post & Pre & Post & Pre & Post \\ \hline
Ev101 & Birthday Party & 254 & 237 & 4.3\% & 2.1\% & 25.2\% & 19.8\% & 23.0\% & 27.9\% \\ \hline
Ev102 & Flash Mob & 157 & 150 & 8.9\% & 4.7\% & 25.5\% & 22.0\% & 29.5\% & 33.0\% \\ \hline
Ev103 & Gtg Vehicle Unstuck & 152 & 141 & 13.8\% & 7.1\% & 39.5\% & 34.8\% & 19.8\% & 25.3\% \\ \hline
Ev104 & Parade & 254 & 230 & 4.7\% & 2.2\% & 28.7\% & 21.3\% & 19.1\% & 23.4\% \\ \hline
Ev105 & Atmptg Board Trick & 311 & 194 & 20.3\% & 2.6\% & 53.7\% & 25.8\% & 7.0\% & 22.8\% \\ \hline
Ev106 & P Grooming Animal & 139 & 139 & 3.6\% & 3.6\% & 23.7\% & 23.7\% & 35.8\% & 35.8\% \\ \hline
Ev107 & Pers Feeding Animal & 229 & 220 & 4.4\% & 2.3\% & 21.4\% & 18.2\% & 29.9\% & 32.9\% \\ \hline
Ev108 & Pers Landing a Fish & 148 & 143 & 10.1\% & 7.0\% & 40.5\% & 38.5\% & 19.8\% & 20.9\% \\ \hline
Ev109 & Wedding Ceremony & 223 & 219 & 3.6\% & 2.3\% & 21.5\% & 20.1\% & 25.7\% & 26.4\% \\ \hline
Ev110 & Woodworking Proj & 155 & 150 & 7.7\% & 4.7\% & 35.7\% & 31.3\% & 31.0\% & 33.6\% \\ \hline
\end{tabular}}
\caption{Number of positive videos for each event in the draft (pre-pruning) and final (post-pruning) corpus indexes, and distribution of positive videos per user in each version.} 
\label{tab:usersplit-comp}
\end{table*}

The differences between events in user distribution (before or after pruning) do not necessarily reflect anything about those particular events, or even about how those events are generally represented in the YFCC100M dataset---especially as, where the YFCC100M dataset had an abundance of videos for an event of which we collected only a portion (see Table~\ref{tab:excludedterms} below), the distribution of videos-per-user merely reflects which videos and users the video collectors happened upon first.\fn{The events where the earliest draft of the corpus index was skewed towards particular users included both events where the corpus collectors had needed to use only a small proportion of the available videos and events where they felt they had ``tapped out'' the supply of eligible videos. However, it is worth noting that the latter type of video had a narrower range of average videos per user, from 1.3 to 1.7 (which might still present a problem in a sufficiently small event set), while for events with a larger overall body of available videos, there was wider variation in average videos per user in that preliminary list, ranging from 1.4 to 3.1.} It would require a more thorough review of the entire YFCC100M dataset to determine whether it in fact the case that, for example, the skating videos are uploaded mainly by a small group of prolific enthusiasts while many people have a few birthday videos to share. However, the differences in user distribution curves within YLI-MED do reflect the general fact that events differ with respect to how people want to share records of them---and that, where there are fewer videos for an event, sparsity of data may accentuate user skew---which is a challenge that video classification and retrieval systems need to be able to work around (see \S\ref{sec:extratrick}).

\subsubsection{Splitting into Training and Test Sets}
\label{sec:splitting}

Once the candidate video index had been verified, cleaned up, and pruned, we split each set of positive-instance videos into standard Training and Test sets, so that researchers can compare results. We do not provide a separate ``development'' set, nor have we held back any data for future evaluations.  

We grouped the videos for each event category by user, then selected random users to make up training sets of 100 videos each; the remainder of the videos for each event became the test sets (ranging from 39 to 137 events).\fn{The released corpus index includes a unique identifier for each user.} 
Selection of users to go in each set was random within the following constraints:
\bi
\item In cases where there were less than 200 videos for an event (i.e., where the test set would be fewer than 100 videos), we constrained the selection process so that the test set would not have more than 5\% of its videos being from a single user. This meant that, in such cases, the more prolific users were assigned to the training set. 
\item For {\it Person Attempting a Board Trick}, we also constrained the process so that tricks using the three different subtypes of board (skate, snow, and surf) would be represented in approximately the same proportion (6:3:1). 
\ei

\noindent The separation of users into training vs.\ test was performed only within each event set, so, for example, a user whose contributions were assigned to the training set for Ev104 might have contributions in the test set for Ev106.

Table~\ref{tab:usersplit} shows the number of users assigned to the standard training set and the standard test set for each event, along with the maximum percentage of videos contributed by any given user (which may be higher or lower than the maximum percentage for all of the positive videos for that event taken together, represented in Table~\ref{tab:vidsperuser}).

\begin{table*}[htp]
\centering
{\renewcommand{\arraystretch}{1.2}
\begin{tabular}{ | p{0.5in} | p{2.25in} || p{0.65in} | p{0.65in} || p{0.65in} | p{0.65in} | }
\hline
\multirow{2}{*}{Event \#} & \multirow{2}{*}{Event Name} & \multicolumn{2}{|c||}{Number of Users} & \multicolumn{2}{c|}{Maximum \% per User} \\ \cline{3-6}
 & & Training & Test & Training & Test \\ \hline
Ev101 & Birthday Party & 70 & 95 & 5.0\% & 3.7\% \\ \hline
Ev102 & Flash Mob & 69 & 43 & 7.0\% & 4.0\% \\ \hline
Ev103 & Getting a Vehicle Unstuck & 59 & 36 & 10.0\% & 4.9\% \\ \hline
Ev104 & Parade & 62 & 79 & 5.0\% & 3.8\% \\ \hline
Ev105 & Person Attempting a Board Trick & 51 & 66 & 5.0\% & 4.3\% \\ \hline
Ev106 & Person Grooming an Animal & 72 & 37 & 5.0\% & 5.1\% \\ \hline
Ev107 & Person Hand-Feeding an Animal & 74 & 90 & 5.0\% & 4.2\% \\ \hline
Ev108 & Person Landing a Fish & 49 & 37 & 10.0\% & 4.7\% \\ \hline
Ev109 & Wedding Ceremony & 69 & 79 & 5.0\% & 4.2\% \\ \hline
Ev110 & Working on a Woodworking Project & 71 & 42 & 7.0\% & 4.0\% \\ \hline
\end{tabular}}
\caption{Number of users contributing to the Training and Test sets of positive videos for each event, and maximum contribution from any user.}
\label{tab:usersplit}
\end{table*}

The near misses and related videos were assigned to a separate set, the Supplemental set, as different researchers might wish to use them in different ways.
Some of the videos in the Supplemental set were contributed by the same users that contributed to either the training set or the test set for that event, and no effort was made to constrain the distribution of videos per user.

\subsubsection{An Opportunity for Experimentation with User Bias}
\label{sec:extratrick}

Although there is a general view in the multimedia retrieval community that training overfit and test skew may result if too great a proportion of the dataset is supplied by a single user, and some evidence from particular use cases like those mentioned in \S\ref{sec:userbias}, the degree of the potential problems and the general parameters determining how they might affect classification and retrieval results are not yet well-studied.

For researchers interested in exploring this territory and how best to build robust systems that can deal with such skew, we are releasing a separate alternative training pool for Ev105 {\it Person Attempting a Board Trick} (Training-Alt), that includes both the 100 training videos from the main (standard) dataset and the 111 videos that were excluded from the index.\fn{The index file is {\it YLI-MED\_v.X\_Ev105\_Alternate\_Training.txt}.} (The videos in the main test set are {\it not} included in the alternate Ev105 training index.) These 211 videos were contributed by 48 users, with an uneven distribution of videos per user. The maximum number of videos per user is 63, or 30.0\% of the alternate training pool altogether. The average number of videos per user is 4.4, or 2.1\% of the full alternate training set (as compared with an average of 2.0 videos per user (2.0\%) for the main Ev105 training set). The top 10 contributors collectively contributed 78.7\% of the alternate training set.

We invite researchers interested in exploring overfit and skew to experiment with creating alternate training sets for Ev105 out of the videos in the alternate index. The alternate index includes the annotations for the subcategories of board trick being attempted (skate, snow, or surf); researchers should be aware that the distribution of videos per user patterns differently for the different subcategories, with the most prolific users mostly contributing skateboarding videos. To construct alternative training set(s) that are comparable to the main training set and to the test set in terms of subcategory distribution, the ratio of Skate:Snow:Surf videos should be roughly 6:3:1.\fn{This distribution is an effect of the corpus-collection process and should not be taken as indicative of ``natural'' distribution in the larger dataset.}

\subsubsection{A Note on Near Misses and Related Videos}
\label{sec:nonpos}

Near misses and related videos were collected during the course of collecting the positive videos; however, although the collectors were instructed to index all near misses and related videos they found, in practice, indexing was very spotty. The majority of the near misses and related videos in the final corpus index are in fact videos that the collectors had judged to be positive instances, but the verifiers did not. A few were also gathered in the process of making the ``negative'' video set (Ev100 {\it None of the Above}); see \S\ref{sec:spotcheck}. 

In sum, the collection process for near misses and related videos was not particularly systematic, and coverage varies widely across the events---between 2 and 33 near misses per event and between 3 and 11 related videos per event. In addition, as we noted in \S\ref{sec:splitting}, no effort was made to constrain the distribution of videos per user. 

For all of these reasons, the near misses and related videos (the Supplemental set) should be considered ``use at your own risk'', and should in no way be considered a representative sample of evidentiarily similar videos nor of the related conceptual space for a given event.\fn{Similarly, the fact that there are nearly twice as many near misses in the index as related videos is an effect of the relative frequency of demotion from positives to near misses in verification, rather than indicating anything about the nature of the source corpus.}

\subsection{Collection and Verification of Negative Videos}
\label{sec:negatives}

For the most part, the ``negative'' videos (Ev100 {\it None of the Above}), i.e., those that do not depict any of the target events, constitute a ``low-confidence'' unconfirmed negative set, in that 99\% of them have not been individually viewed to confirm that they are not instances of any of the target events.\fn{This differentiates YLI from the HAVIC data used for TRECVID MED, where each video designated ``negative'' had been individually screened (\citealt{HAVIC,TV2011}). Screening 50,000 videos would have been prohibitively time-consuming for this initial corpus-building effort; however, we hope to increase the number of screened videos in later releases. Some possible consequences of this methodology are discussed in \S\ref{sec:biasesdisc}.}  However, other measures were taken to reduce the likelihood that they might be positive examples; for example, the potential negative videos, gathered by random selection from among all of the videos in the YFCC100M dataset, were screened for textual metadata that might indicate they depicted one of the target events. About 1\% of the videos designated {\it None of the Above} were checked by a human annotator to confirm that they were not actually examples of any of the target events. These processes are laid out in detail in the remainder of this section.

It should be noted that, because most of the videos were not individually reviewed, we did not exclude videos that did not meet the criteria described in \S\ref{sec:collection} for fully reviewed videos, such as having an audio track or being at least 3 seconds long. Around 4\% of the negatives in the final corpus were thus shorter than 3 seconds.

\subsubsection{Selection and Metadata-Based Exclusion}
\label{sec:exclusion}

We began by randomly selecting \~{}50,000 videos from the YFCC100M dataset of \~{}800,000 videos. After checking this against the index of fully reviewed videos and removing any we had collected directly from Flickr, we had 49,870 potential videos to include in the negatives index.

We next excluded any videos that had words or strings in their titles, descriptions, or tags that had a reasonable likelihood of indicating they might depict the target events.\fn{The YFCC100M metadata does not include all text description, only titles, tags, and descriptions. Album titles, user comments, and so forth are not included, so could not be used for this purpose.} For half of the target events, the corpus-collectors had already gathered all of the videos tagged or described with any of the most obvious words in English and other languages frequently used on Flickr (and in some cases, had tried many non-obvious ones as well), i.e., we had probably come close to tapping out the YFCC100M dataset with respect to videos of those events that could be found using textual metadata. For the other half of the events, the corpus-collectors had {\it not} needed to screen all of the videos with potentially relevant words in the metadata, because they had already collected enough to meet our numerical goals. For these latter events, where we considered it likely that the randomly selected potential negatives might still have contained positive videos, we removed from the ``negative'' event set any videos containing in their textual metadata any of the most fruitful search terms we had used to find videos for that event (or very similar search terms), along with some additional obvious search terms in languages the collectors had not explored. Table~\ref{tab:excludedterms} shows the terms we used to automatically exclude videos from the negatives.\fn{Table~\ref{tab:excludedterms} does not list strings (mostly non-English) we tried that were not found in the metadata for any of the potential negatives. Strings that were used to exclude additional negative videos after the release of YLI-MED Version 1.0 are not listed.}

\begin{table*}[htp]
\centering
{\renewcommand{\arraystretch}{1.2}
\begin{tabular}{| p{2in} | p{2.8in} | p{1in} |}
\hline
Event & Strings Excluded & \# of Videos with Those Strings \\ \hline
Birthday Party & anniversaire, birthday, complea, geburtstag & 339 \\ \hline
Parade & parada, parade & 192 \\ \hline
Person Attempting a Board Trick & sk8, skateboard, skating, snowboard, surf; ollie & 295 \\ \hline
Person Hand-Feeding an Animal & fed, feed & 180 \\ \hline
Wedding Ceremony & hochzeit, mariage, marriage, wedding & 213 \\ \hline
\end{tabular}}
\caption{Events with many more examples in YFCC100M than we needed, and strings used to automatically exclude videos of those events from the Ev100 {\it None of the Above} set (at any stage; semicolon indicates videos removed after random verification round).}
\label{tab:excludedterms}
\end{table*}

We spot-checked some of the excluded videos and determined that around 25\%--30\% were likely actually examples of the event, evidential near misses, or closely related. However, there was a great deal of variation between the excluded strings (from 5\%--100\%), depending on the number of instances of the string, how discriminatory the word was (in TF/IDF terms), whether the word was polysemous or otherwise likely to frequently be used in context unrelated to the event, and to what degree the corpus-collectors had already reviewed the videos with that string in the metadata. In all, 1,214 videos were excluded from the negatives in this fashion (including a few removed after the random verification process described in \S\ref{sec:spotcheck}.\fn{This figure does not match the sum of the figures in Table~\ref{tab:excludedterms} because some videos had terms potentially related to more than one event.}

Using such a broad method means that it is of course possible that we removed videos that did not depict the event but that were related to it closely enough that they would be a good test of discrimination. However, this function can be served by the Near Miss and Related videos we indexed while finding the positive videos using those search terms---and in any case, if there were any such videos, they should properly have been labeled Related rather than Negative.

\subsubsection{Creating the ``Confirmed'' Negative Set}
\label{sec:spotcheck}

Finally, we selected 500 random videos from the Ev100 (negative) index---about 1\%---to be reviewed by at least one of the corpus-builders. Within the 498 videos that were viewable at the time of review, the verifiers found 7 videos that they judged to be positive examples of our target events (1.4\% of the reviewed negatives), as well as 4 near misses (0.8\%) and 4 related videos (0.8\%). These videos were excluded (or added to the event sets) and the remainder were designated as confirmed ``high-confidence'' negatives (recorded in the corpus file under the column `Neg Confirmation'). 

Since each video was being viewed by only one verifier, to validate the results, we included some known-positive videos (never before seen by the corpus-builders) in the sets of negative videos to be reviewed. The verifiers caught 90\% of these.\fn{We checked back with the verifiers and confirmed that omissions were indeed attentional blips or record-keeping errors, not differences in subjective judgment.} This 10\% error rate suggests that the proportion of positives in the reviewed negatives may have been as high as 1.6\%.
Assuming that the (unreviewed) videos categorized as unconfirmed negatives have approximately the same likelihood of actually being positive instances, researchers should thus assume that accuracy measurements for classification or detection algorithms using this dataset have an error margin of up to 2\%.

To some degree, the review of random videos served as a test of how well the method of automatically excluding videos based on textual metadata worked, given that the metadata for videos does not necessarily describe what happens in them, and many videos do not have textual metadata at all (\citealt{ulges2010visual}). Of the positives, near misses, and related videos found among the potential negatives, the vast majority (100\% of the positives and 63\% of the near misses and related videos) belonged to the event categories for which video-collecting had been relatively easy (those listed in Table~\ref{tab:excludedterms}), i.e., where we had reached our target number without screening all of the videos retrieved using obvious search terms. Looking at the textual metadata for the 15 videos we found, we found that most did not have tags or text descriptions. The remainder either did not have highly informative tags/descriptions/titles or had tags/descriptions/titles that were related to the target event only indirectly, or that were not distinctive enough to have merited automatically removing all such videos from the negative set (e.g., {\it festival} 
for Ev104 {\it Parade} or {\it Mr. and Mrs. X} for Ev109 {\it Wedding Ceremony}). We thus only found one additional string that needed to be excluded (see Table~\ref{tab:excludedterms}).

\subsubsection{Division into Training and Test Sets}
\label{sec:negdiv}

After the procedures described in the previous subsubsections, we had a total of 48,638 negative videos in the Ev100 ``event'' set, including 48,155 unconfirmed and 483 confirmed negatives.\fn{The figures in this section are for Version 1.0 of the corpus.} This provides a negative:positive ratio of 27:1 for the YLI-MED corpus---higher than that of most similar corpora, but leaving room for us to add additional events without needing to also add additional negative videos. We divided these into a Training set of 5,000 videos and a Test set of 43,138 videos. 

As with the videos for each event set, we grouped the videos by user and then randomly assigned users to either the Training set or the Test set, with the only constraint being to ensure that the relatively smaller Training set did not have more than 2\% of its videos from any one user.\fn{No pruning was necessary to produce a reasonable distribution of videos per user for the negatives index as a whole, given its size.} As with the positive event sets, no effort was made to ensure that assignment of users to Training or Test was consistent with the other event sets (only within Ev100).

After division, the Training set for Ev100 contained videos from 1641 users and the Test set contained videos from 14,671 users. The Training set had a maximum of 1.96\% of its videos contributed by any one user, while the Test set had a maximum of 1.85\% from any one user.

The Training and Test sets each included 1.0\% confirmed and 99.0\% unconfirmed negative videos.

\subsubsection{Other Negative Confirmation Mechanisms}
\label{sec:negbackstop}

As we begin to use this dataset in research, we will no doubt run experiments in which our classification and detection systems produce apparent false positive results. Such experiments will provide an opportunity for further refinement of the negative videos, if any of the ``false'' positives actually are positive instances of the event.

Finally, we hope to receive feedback from the community of other researchers using this corpus. If, in the course of your research, you discover any positive instances of the target events among the videos labeled Negative (in the Ev100 event set), please email \ylicontact so we can remove them from the set.

\section{Characteristics of the YLI-MED Corpus}
\label{sec:character}

This section provides extensive descriptive statistics about YLI-MED Version 1. (Except as otherwise noted, statistics refer specifically to Version 1.0.) First, in \S\ref{sec:charovw}, we provide a general overview of the corpus, including videos per event and levels of annotator agreement and confidence about those videos. \S\ref{sec:peruser} provides  information about the distribution of videos per user. In \S\ref{sec:nonevent}, we give an overview of the additional corpus characteristics we annotated (besides events), including post-production characteristics and language. In \S\ref{sec:ttcomp}, we compare the Training and Test sets to identify any statistically significant differences that could affect classification and detection performance, and in \S\ref{sec:psych}, we describe preliminary findings about how post-production characteristics affect annotators' event judgments.

\subsection{Overview of Characteristics for YLI-MED v.1.0}
\label{sec:charovw}

YLI-MED Version 1.0 contains 50,638 videos, including 1823 videos judged to be positive instances of event categories, 115 evidential near misses, and 62 videos related to the target events, as well as 48,638 negative videos, i.e., videos that do not depict any of the target events, nor are they closely related to them.\fn{Versions 1.1 and later of YLI-MED have slightly fewer videos than Version 1.0, as videos that were removed from Flickr have been subtracted from the index.} 

\subsubsection{Event Breakdown}

Table~\ref{tab:counts} shows the final numbers of videos for each event category in the released corpus (including positives, near misses, and related videos). There is a fair amount of variation in the number of videos for each event, from 139 positive examples for Ev106 {\it Person Grooming an Animal} to 237 positive examples for Ev101 {\it Birthday Party}.

\begin{table*}[htp]
\centering
{\renewcommand{\arraystretch}{1.2}
\begin{tabular}{| p{0.5in} | p{2.25in} | p{0.9in} | p{0.9in} | p{0.9in} |}
\hline
Event \# & Event Name & Positive Instances & Near Misses & Related Videos \\ \hline
Ev101 & Birthday Party & 237 & 6 & 5 \\ \hline
Ev102 & Flash Mob & 150 & 33 & 4 \\ \hline
Ev103 & Getting a Vehicle Unstuck & 141 & 24 & 5 \\ \hline
Ev104 & Parade & 230 & 4 & 4 \\ \hline
Ev105 & Person Attempting a Board Trick & 194 & 16 & 9 \\ \hline
Ev106 & Person Grooming an Animal & 139 & 5 & 6 \\ \hline
Ev107 & Person Hand-Feeding an Animal & 220 & 2 & 3 \\ \hline
Ev108 & Person Landing a Fish & 143 & 15 & 9 \\ \hline
Ev109 & Wedding Ceremony & 219 & 2 & 6 \\ \hline
Ev110 & Working on a Woodworking Project & 150 & 8 & 11 \\ \hline
 & Total & 1823 & 115 & 62 \\ \hline
\end{tabular}}
\caption{Number of videos for each event in YLI-MED Version 1.0.}
\label{tab:counts}
\end{table*}

Each event set includes a standard set of 100 positive Training videos; the remainder of the videos for the event are in a standard Test set. (The near misses and related videos are in a separate Supplemental set.) The negative videos are divided into 5,000 Training videos and 48,138 Test videos.

\subsubsection{Agreement and Annotator Confidence}

The great majority of the ``positive'' videos were judged as positive instances of their event category by three out of the three annotators that viewed them (the collector and two verifiers), while the great majority of near misses and related videos were judged so by only two out of the three annotators (with the third annotator assigning a different relation).\fn{The full procedures for calculating agreement and confidence are described in \S\ref{sec:iaa}.} Table~\ref{tab:agrbyrel-final} shows the proportion of videos with full vs.\ partial agreement for each type of relation to the target event, and Table~\ref{tab:relbyagr-final} shows how the degree of agreement varies by relation.\fn{Agreement levels for near misses and related videos are the same in Table~\ref{tab:agrbyrel-final} as in Table~\ref{tab:agrbyrel-init} because no near misses nor related videos were removed from the initial index to create the released version.} 

\begin{table*}[htp]
\centering
{\renewcommand{\arraystretch}{1.2}
\begin{tabular}{ | p{1.4375in} | p{1.4375in} | p{1.4375in} | p{1.4375in} | }
\hline
Agreement Level & Positive Instances & Near Misses & Related Videos \\ \hline
Partial Agreement & 7.3\% & 82.6\% & 91.9\% \\ \hline
Full Agreement & 92.7\% & 17.4\% & 8.1\% \\ \hline
\end{tabular}}
\caption{Level of agreement for each type of relation to event category among reviewed videos in the released corpus. (Partial = 2/3 annotators agreed; full = 3/3 annotators agreed.)} 
\label{tab:agrbyrel-final}
\end{table*}

\begin{table*}[htp]
\centering
{\renewcommand{\arraystretch}{1.2}
\begin{tabular}{ | p{1.4375in} | p{1.4375in} | p{1.4375in} | p{1.4375in} | }
\hline
Agreement Level & Agreement as \% of Reviewed Videos & Consensus Relation to Event & Relation as \% of Agreement Level \\ \hline\hline
\multirow{3}{*}{Partial Agreement} & \multirow{3}{*}{14.3\%} & Positive & 46.7\% \\ \cline{3-4}
 & & Near Miss & 33.3\% \\ \cline{3-4}
 & & Related & 20.0\% \\ \hline\hline
\multirow{3}{*}{Full Agreement} & \multirow{3}{*}{85.8\%} & Positive & 98.5\% \\ \cline{3-4}
& & Near Miss & 1.2\% \\ \cline{3-4}
 & & Related & 0.3\% \\ \hline
\end{tabular}}
\caption{Percentage of reviewed videos in the released corpus achieving each level of agreement, broken down by the consensus relation to the target event.}
\label{tab:relbyagr-final}
\end{table*}

Table~\ref{tab:agrbyevent-final} breaks down the agreement levels by event category, showing wide variation between the categories, from {\it Flash Mob}, with 79.3\% of the positive videos getting full agreement as to their positivity, to {\it Parade}, with 98.3\%. (Agreement differences between events are statistically significant both across the positive videos and across all of the reviewed videos; p $<$ 0.0001, chi-square.)

\begin{table*}[htp]
\centering
{\renewcommand{\arraystretch}{1.2}
\begin{tabular}{ | p{0.5in} | p{2.25in} || p{0.65in} | p{0.65in} || p{0.65in} | p{0.65in} | }
\hline
\multirow{3}{*}{Event \#} & \multirow{3}{*}{Event Name} & \multicolumn{2}{|c||}{Agreement:} & \multicolumn{2}{c|}{Agreement:} \\ 
 & & \multicolumn{2}{|c||}{Positive Videos} & \multicolumn{2}{c|}{All Reviewed Videos } \\ \cline{3-6}
 & & Partial & Full & Partial & Full \\ \hline
Ev101 & Birthday Party & 4.2\% & 95.8\% & 6.9\% & 93.1\% \\ \hline
Ev102 & Flash Mob & 20.7\% & 79.3\% & 36.4\% & 63.6\% \\ \hline
Ev103 & Getting a Vehicle Unstuck & 6.4\% & 93.6\% & 21.8\% & 78.2\% \\ \hline
Ev104 & Parade & 1.7\% & 98.3\% & 4.6\% & 95.4\% \\ \hline
Ev105 & Person Attempting a Board Trick & 6.2\% & 93.8\% & 13.7\% & 86.3\% \\ \hline
Ev106 & Person Grooming an Animal & 12.2\% & 87.8\% & 17.3\% & 82.7\% \\ \hline
Ev107 & Person Hand-Feeding an Animal & 4.5\% & 95.5\% & 6.7\% & 93.3\% \\ \hline
Ev108 & Person Landing a Fish & 14.0\% & 86.0\% & 23.4\% & 76.6\% \\ \hline
Ev109 & Wedding Ceremony & 2.3\% & 97.7\% & 5.7\% & 94.3\% \\ \hline
Ev110 & Working on a Woodworking Project & 10.0\% & 90.0\% & 17.2\% & 82.8\% \\ \hline
\end{tabular}}
\caption{Level of agreement per event for positive videos and for all reviewed videos in the released corpus (positive, near miss, and related).}
\label{tab:agrbyevent-final}
\end{table*}

Table~\ref{tab:agrconf} shows the average confidence for the positive videos in each event category, and the relationship between confidence and annotator agreement on whether the videos were in fact positive examples.\fn{Confidence scores for each video were calculated only among those annotators who judged it positive, on a scale from 1 to 3.} The average confidence across all positive videos in every category is 1.95 for videos on which agreement was only partial and 2.77 for videos on which all three annotators agreed they were positive examples; this difference is statistically significant at p $<$ 0.0001 (Student's unpaired t-test, two-tail). Confidence was also more variable for events with only partial agreement, with a standard deviation of 0.68 across all videos with partial agreement and a standard deviation of 0.48 across videos with full agreement. 

\begin{table*}[htp]
\centering
{\renewcommand{\arraystretch}{1.2}
\begin{tabular}{ | p{0.5in} | p{2in} || p{0.525in} | p{0.525in} || p{0.525in} | p{0.525in} || p{0.525in} | }
\hline
Event \# & Event Name & \% with Partial Agreement & Avg. Confidence for Partial & \% with Full Agreement & Avg. Confidence for Full & Total Avg. Confidence \\ \hline
Ev101 & Birthday Party & 4.2\% & 2.20 & 95.8\% & 2.91 & 2.88 \\ \hline
Ev102 & Flash Mob & 20.7\% & 1.95 & 79.3\% & 2.44 & 2.34 \\ \hline
Ev103 & Getting a Vehicle Unstuck & 6.4\% & 2.06 & 93.6\% & 2.56 & 2.53 \\ \hline
Ev104 & Parade & 1.7\% & 2.25 & 98.3\% & 2.84 & 2.83 \\ \hline
Ev105 & Person Attempting a Board Trick & 6.2\% & 2.29 & 93.8\% & 2.74 & 2.71 \\ \hline
Ev106 & Person Grooming an Animal & 12.2\% & 1.94 & 87.8\% & 2.87 & 2.76 \\ \hline
Ev107 & Person Hand-Feeding an Animal & 4.5\% & 2.25 & 95.5\% & 2.90 & 2.87 \\ \hline
Ev108 & Person Landing a Fish & 14.0\% & 1.60 & 86.0\% & 2.71 & 2.55 \\ \hline
Ev109 & Wedding Ceremony & 2.3\% & 2.00 & 97.7\% & 2.88 & 2.86 \\ \hline
Ev110 & Wkg on a Woodworking Project & 10.0\% & 1.67 & 90.0\% & 2.51 & 2.42 \\ \hline
& Total: Positives & 7.3\% & 1.95 & 92.7\% & 2.77 & 2.71 \\ \hline
\end{tabular}}
\caption{Average annotator confidence in positive judgments per event category, broken down by partial vs.\ full agreement on videos' relation to the event.}
\label{tab:agrconf}
\end{table*}

As we noted in \S\ref{sec:iaa}, the significant variation in agreement and confidence across events may stem from several sources. Some potential sources of disagreement and lack of confidence in the evidence include:

\bi
\item The event itself may be a contested category among speakers of English in general (for example, there is much discussion in Wikipedia and other sources of what counts as a ``real'' {\it flash mob}), or it may involve participants or products that are contested categories (for example, {\it trick}).
\item The subjective judgments of the specific annotators may have disagreed about what counts as belonging to an event category or an aspect thereof, even if the category isn't frequently and publicly contested (for example, there were discussions between the corpus-builders and the staff researchers about what counts as {\it grooming} or a {\it project}).
\item The event as formulated by the YLI-MED research team (based on the TRECVID MED event names (\citealt{TV2011})) may be a subcategory or supercategory of a more basic-level event category (\citealt{rosch1978}) (for example, {\it catching a fish} may be a more basic-level event category than the subevent {\it landing a fish}).
\item How clear the observable evidence tends to be for a particular type of video vs.\ how much inference is required on the part of the viewer. All of the target events involve specific observable actions by the protagonists, but for some events, it can more often be the case that a video could lack evidence that shows {\it why} they are doing what they are doing (for example, annotators mentioned having difficulty with {\it Getting a Vehicle Unstuck} when they could not observe the mechanical details directly enough to tell whether the vehicle was actually stuck or whether it was engine failure; similarly, it was often unclear whether someone quickly reeling in a fishing line actually had a fish on the other end).
\item Events vary in the degree to which they are defined by the intentions of the protagonists (for example, there was discussion between the corpus-builders and the staff research team about whether sheep-shearing should count as {\it Grooming an Animal} given that, while it is necessary for the animal's hygiene, it is performed mainly for the sake of producing wool).
\ei

\noindent The event definitions and decision rubrics in  Appendix~\ref{sec:defsapp} provided some guidance on these points, but of course, it would not be possible for them to cover every potential source of disagreement about how to categorize a video---nor would that necessarily be desirable.

\subsection{Distribution of Videos per User}
\label{sec:peruser}

As described in \S\ref{sec:userbias} and \S\ref{sec:splitting}, we endeavored to ensure that the videos in both the standard training and the test sets were contributed by a variety of different users.\fn{By {\it user} or {\it uploader}, we mean a single Flickr account; of course, some individual users may have had multiple accounts with different user IDs.} Table~\ref{tab:vidsperuser} shows the average and maximum numbers of positive-instance videos contributed by each user for each event as well as the negative set (Ev100), and the average and maximum percentage of each event set that was contributed by a single user.\fn{Some numbers are repeated from Table~\ref{tab:usersplit-comp}. See \S\ref{sec:splitting} and \S\ref{sec:negdiv} for a breakdown into Training and Test sets.} In addition, it shows the average and maximum numbers and percentages for all of the positives together (where some users may have contributed videos to multiple events) and for the YLI-MED corpus as a whole (including the positives and negatives and the Supplemental set of near misses and related videos), as well as the same statistics for the whole Yahoo Flickr Creative Commons 100 Million dataset (\citealt{YFCC,YFCC2014}) from which the YLI-MED corpus is drawn.\fn{Videos-per-user statistics for the alternate user-skewed training set for Ev105 are given in \S\ref{sec:extratrick}.} 

\begin{table*}[htp]
\centering
{\renewcommand{\arraystretch}{1.2}
\begin{tabular}{ | p{0.35in} | p{1.75in} || p{0.4in} | p{0.4in} || p{0.4in} | p{0.4in} | p{0.4in} | p{0.4in} | p{0.4in} |}
\hline
Event \# & \multirow{2}{*}{Event Name} & \# of Videos & \# of Users & \multicolumn{2}{l|}{Max Vids/User} & \multicolumn{2}{l|}{Avg. Vids/User} & Top 10 \\ \cline{5-8}
 & & & & \# & \% & \# & \% & Users \\ \hline\hline
Ev101 & Birthday Party & 237 & 165 & 5 & 2.1\% & 1.44 & 0.6\% & 19.8\% \\ \hline
Ev102 & Flash Mob & 150 & 112 & 7 & 4.7\% & 1.34 & 0.9\% & 22.0\% \\ \hline
Ev103 & Getting a Vehicle Unstuck & 141 & 95 & 10 & 7.1\% & 1.48 & 1.1\% & 34.8\% \\ \hline
Ev104 & Parade & 230 & 141 & 5 & 2.2\% & 1.63 & 0.7\% & 21.3\% \\ \hline
Ev105 & Pers Attempting Board Trick & 194 & 114 & 5 & 2.6\% & 1.70 & 0.9\% & 25.8\% \\ \hline
Ev106 & Person Grooming an Animal & 139 & 109 & 5 & 3.6\% & 1.28 & 0.9\% & 23.7\% \\ \hline
Ev107 & Person Hand-Feeding Animal & 220 & 164 & 5 & 2.3\% & 1.34 & 0.6\% & 18.2\% \\ \hline
Ev108 & Person Landing a Fish & 143 & 86 & 10 & 7.0\% & 1.66 & 1.2\% & 38.5\% \\ \hline
Ev109 & Wedding Ceremony & 219 & 148 & 5 & 2.3\% & 1.48 & 0.7\% & 20.1\% \\ \hline
Ev110 & Wkg on a Woodworking Proj & 150 & 113 & 7 & 4.7\% & 1.33 & 0.9\% & 31.3\% \\ \hline\hline
 & All Positives (across events) & 1823 & 1133 & 12 & 0.7\% & 1.61 & 0.1\% & 5.6\% \\ \hline
Ev100 & None of the Above (negatives) & 48,638 & 16,311 & 809 & 1.7\% & 2.98 & 0.006\% & 5.7\% \\ \hline
 & All YLI-MED (positive, negative, and Supplemental) & 50,638 & 16,729 & 809 & 1.6\% & 3.03 & 0.006\% & 5.5\% \\ \hline
 & All YFCC100M (source dataset) & 793,436 & 57,712 & 13,298 & 1.7\% & 13.75 & 0.002\% & 5.8\% \\ \hline
\end{tabular}}
\caption{Distribution of videos per user among positive videos by event set and among various combined groupings.}
\label{tab:vidsperuser}
\end{table*}

There is a fair amount of variation between the (relatively small) individual event sets; most tellingly, as we noted in \S\ref{sec:userbias}, the percentage of videos contributed by the ten most prolific users varies between 18\% for Ev107 {\it Person Hand-Feeding an Animal} and 38\% for Ev108 {\it Person Landing a Fish}. The negative set is much closer to being representative of the user distribution across the whole YFCC100M dataset in terms of percentages (as it was a random selection).

As the negatives make up the bulk of YLI-MED, the YLI-MED corpus as a whole is thus fairly representative of the larger YFCC100M dataset in user distribution, in terms of the maximum percent of the videos contributed by a single user (1.6\% for YLI-MED vs.\ 1.7\% for YFCC100M), average percent of videos contributed by a single user (0.006\% vs.\ 0.002\%), and proportion of the videos that were contributed by the top ten users (5.5\% vs.\ 5.8\%). YLI-MED is actually less skewed than YFCC100M as a whole in terms of the distribution of high-contributing users' videos over the bulk of the corpus, in that half of the videos in the YFCC100M were contributed by just 3\% of the users, while for YLI-MED, the top 10\% of the users contributed half of the videos. 

The median number of videos per user is 1, for each of the events as well as for the negative videos and for YLI-MED as a whole. Similarly, the mode videos/user is 1 for all event sets, for the negatives, and for all YLI-MED.

\subsection{Non-Event Characteristics}
\label{sec:nonevent}

This section describes the distribution of positive-example videos in YLI-MED in terms of various post-production characteristics (\S\ref{sec:postprod}) and in terms of language (\S\ref{sec:language}). For information about the geographical and temporal distribution of the YLI-MED positive videos, see the comparison with the overall YFCC100M distributions in \S\ref{sec:procchoices1}.

\subsubsection{Post-Production Characteristics}
\label{sec:postprod}

In addition to determining whether a video belonged to one of the target event categories,  the corpus-builders also annotated the candidate videos with regard to whether they had certain production and post-production characteristics: containing or consisting of animation or CGI, having added text such as title screens or subtitles, and having an added music track. (See 
\S\ref{sec:annot} for details.) Table~\ref{tab:postprodpos} shows the distribution of these features across positive videos for each of the target events.\fn{The final column in Table~\ref{tab:postprodpos}, ``Any Post-Production'', refers {\it only} to the three features we targeted for annotation, and {\it not} any other post-production features such as voiceovers, montaging, audio or visual warping, etc.} It also includes the distribution for all of the positive videos together, and for all of the reviewed videos including positives, near misses, and related videos.\fn{The by-event breakdown for post-production features for all of the reviewed videos combined (including positives, near misses, and related videos) patterns very similarly to the breakdown in Table~\ref{tab:postprodpos} for the positive-instance videos only. Here and in \S\ref{sec:language}, we provide detailed numbers only for the positive videos where patterns are similar, on the assumption that researchers will have more need for specifics as regards the positives. If we have judged incorrectly, you are welcome to contact \ylicontact to obtain the relevant statistics, rather than having to recalculate them yourself.}

\begin{table*}[htp]
\centering
{\renewcommand{\arraystretch}{1.2}
\begin{tabular}{ | p{0.5in} | p{2.25in} | p{0.65in} | p{0.65in} | p{0.65in} | p{0.65in} | }
\hline
Event \# & Event Name & \% with Animation & \% with Added Text & \%  with Musical Score & Any Post-Production \\ \hline
Ev101 & Birthday Party & 1.3\% & 2.5\% & 1.7\% & 3.8\% \\ \hline
Ev102 & Flash Mob & 1.3\% & 3.3\% & 0.7\% & 4.7\% \\ \hline
Ev103 & Getting a Vehicle Unstuck & 1.4\% & 2.1\% & 1.4\% & 2.1\% \\ \hline
Ev104 & Parade & 0.0\% & 2.2\% & 1.3\% & 2.6\% \\ \hline
Ev105 & Person Attempting a Board Trick & 3.6\% & 9.8\% & 13.9\% & 16.0\% \\ \hline
Ev106 & Person Grooming an Animal & 0.0\% & 1.4\% & 4.3\% & 4.3\% \\ \hline
Ev107 & Person Hand-Feeding an Animal & 0.9\% & 1.4\% & 0.9\% & 1.8\% \\ \hline
Ev108 & Person Landing a Fish & 1.4\% & 5.6\% & 2.1\% & 6.3\% \\ \hline
Ev109 & Wedding Ceremony & 0.0\% & 3.2\% & 0.5\% & 3.2\% \\ \hline
Ev110 & Working on a Woodworking Project & 4.7\% & 8.7\% & 10.7\% & 16.7\% \\ \hline
& Total: Positives & 1.4\% & 3.9\% & 3.6\% & 5.9\% \\ \hline
& Total: All Reviewed Videos & 1.3\% & 4.3\% & 3.9\% & 6.5\% \\ \hline
\end{tabular}}
\caption{Selected post-production features for positive-instance videos by event, and for all reviewed videos.}
\label{tab:postprodpos}
\end{table*}

Ev105 {\it Person Attempting a Board Trick} had the highest or second-highest proportion of videos for all of the targeted types of post-production. Overall, board-trick videos in YLI-MED (especially skating videos) are more likely to be artistically refined, due presumably to their function in disseminating culture.\fn{Suggested sources of post-production variation discussed here are speculations based on spot-checks of the videos in the corpus, not facts based on hard numbers.} Their unusually high rate of added music (14\%) also reflects a general tendency (noticed by the research team) for sports videos to be music-tracked, perhaps because the original sound does not provide very much of the interest or information.  However, interpretive and identifying information is often added to sports videos in the form of titles and subtitles---and Ev105 also has the highest rate (10\%) of text added in post-production. (Sports-genre tendencies may also explain the relatively high rate of added text for Ev108 {\it Person Landing a Fish}.)

The other event that stands out in terms of post-production is Ev110 {\it Working on a Woodworking Project}; since these are often how-to videos, it is not surprising that they would have added instructional elements like subtitles and animated  sequences. More surprising is their high rate of musical soundtracking (11\%). In some cases, this may be because they are how-to's, which tend to be more produced. In addition, a number of woodworking videos are in a different genre, where personal project videos are sped up or time-lapsed and then soundtracked with fast-paced music. 

Within those videos that have musical soundtracks, we made further distinctions based on whether the video also retained its original  audio track. Among the positive-instance videos with musical scores, 54\% have the musical score overlaid over the original audio, 9\% have the score interspersed with the original audio, and for 37\%, the music score replaces any original audio entirely. %
The individual events each seem to follow this same pattern (though for events with a very low overall rate of music-tracking, patterns are difficult to discern).

\subsubsection{Language Characteristics}
\label{sec:language}

Annotators recorded information about the language(s) spoken and/or sung in each video and the language(s) of any text added in post-production.\fn{The numbers and statistics given in this section were calculated {\it after} some of the language annotations had been updated and made more consistent in format, after the release of YLI-MED Version 1.0. However, they were calculated {\it before} videos no longer in Flickr had been removed from the corpus, which was done just before the release of Version 1.1. Therefore, the labels referred to in this section match those in Version 1.1 and later, but the numbers of videos correspond with Version 1.0. Note that the problem mentioned in \S\ref{sec:annot}, that annotators may in a few cases have filled in the dominant language of the metadata as the Language Spoken when there was no spoken language, has not yet been fully fixed as of Version 1.2. The statistics about spoken language in this section should thus be considered approximate.} (See \S\ref{sec:annot} for details.)  

The vast majority of positive videos (97\%) have spoken or sung language, while a much smaller percentage (4\%) have added text.\fn{The proportions are quite similar for all of the reviewed videos taken together---including positives, near misses, and related videos---at 96\% with spoken/sung language and 5\% with added text.} Around 3\% of positive videos (36 videos) have both spoken language and added text; Table~\ref{tab:typeoflg} shows the correspondence between the spoken/sung language and the language of added text for the positive videos.\fn{In all tables in this section, ``Added Text'' refers only to text added in post-production, such as titles and subtitles, {\it not} to incidental text appearing in the content of the videos.} In around 80\% of the cases where there is both spoken language (in the video content or added in post-production) and added text, the two languages are the same.\fn{Table~\ref{tab:typeoflg} assumes that if both the spoken language and the added text are marked simply as `Non-English', they are probably the same. Named specific languages and an unidentified `Non-English' language are counted as different languages. For purposes of this table only, videos whose only added text is proper names are counted as having no added text.}

\begin{table*}[htp]
\centering
{\renewcommand{\arraystretch}{1.2}
\begin{tabular}{ | p{2.5in} | p{1in} | p{1in} | p{1.25in} | }
\hline
 & Positive Videos & \multicolumn{2}{|c|}{\multirow{4}{*}{}} \\ \cline{1-2}
Neither (no language) & 2.0\% & \multicolumn{2}{|c|}{} \\ \cline{1-2}
Spoken Language Only &  94.0\% &\multicolumn{2}{|c|}{} \\ \cline{1-2}
Added Text Only & 1.0\% & \multicolumn{2}{|c|}{} \\ \hline
Both Spoken Language and Added Text: & 3.1\% & Out of Total & Out of Vids w/ Both \\ \hline
\ \ \ Same Language(s) &  & 2.5\% & 80.4\% \\
\ \ \ One Same, (at Least) One Different & & 0.2\% & 5.4\% \\
\ \ \ Different Language(s) & & 0.4\% & 14.3\% \\ \hline
\end{tabular}}
\caption{Percentage of positive-instance videos having spoken/sung language, added text, both, or neither.}
\label{tab:typeoflg}
\end{table*}

Table~\ref{tab:alllgs} shows the breakdown of different languages spoken/sung and the breakdown of languages added as text for all of the positive videos.\fn{The annotations found in the corpus index sometimes include ``?'' where the corpus-builder was not fully certain, but language labels with and without ``?'' are counted together for the purposes of tables in this section. Note that videos labeled as `Unintelligible' may or may not have spoken English content (the annotators couldn't even tell that much), while videos labeled as `Non-English' involve a spoken language that is intelligible enough annotators were certain it was not English.} The spoken/sung language in the majority of the videos is English or English and some other language (90\% together), and similar English-dominance is found for added text (89\% together). As we noted in \S\ref{sec:collection}, this is to some degree an effect of the fact that all of the corpus builders spoke English natively or near-natively and so concentrated on English search terms when finding videos---though, as we note below, even for events where they did extensive searching in many other languages, most of the results were in English. 

\begin{table*}[htp]
\centering
{\renewcommand{\arraystretch}{1.2}
\begin{tabular}{| p{2.5in} | p{1.5in} | p{1.5in} |}
\hline
Language & Spoken Language & Language of Added Text \\ \hline
Dutch &  0.1\% &  0.0\% \\ \hline
English &  89.5\% &  88.0\% \\ \hline
French &  0.3\% &  0.0\% \\ \hline
German &  0.1\% &  2.7\% \\ \hline
Italian &  0.3\% &  0.0\% \\ \hline
Japanese &  0.1\% &  1.3\% \\ \hline
Korean &  0.3\% &  0.0\% \\ \hline
Latvian &  0.0\% &  1.3\% \\ \hline
Mandarin &  0.2\% &  0.0\% \\ \hline
Polish &  0.1\% &  0.0\% \\ \hline
Portuguese &  0.1\% &  1.3\% \\ \hline
Spanish &  1.1\% &  1.3\% \\ \hline
Non-English (Unidentified) &  7.0\% &  1.3\% \\ \hline
Danish \& English &  0.1\% &  0.0\% \\ \hline
English \& Italian &  0.1\% &  1.3\% \\ \hline
English \& Spanish &  0.2\% &  0.0\% \\ \hline
English \& Non-English (Unidentified) &  0.3\% &  0.0\% \\ \hline
English \& Unintelligible &  0.1\% &  N/A \\ \hline
Unintelligible &  0.2\% & N/A \\ \hline
Proper Names Only &  N/A & 1.3\% \\ \hline
\end{tabular}}
\caption{Languages spoken/sung, as a percentage of positive videos with spoken/sung language (N=1769), or used for text added in post-production, as a percentage of positive videos with added text (N=75).}
\label{tab:alllgs}
\end{table*}

Table~\ref{tab:spokenbyevent} shows what percentage of videos have spoken or sung language broken down by event. For most events, the rate of spoken language is similar, between 95\% and 99\%; some notable exceptions are Ev105 {\it Person Attempting a Board Trick} and Ev110 {\it Working on a Woodworking Project}, both of which are at 88\% for spoken or sung language (a significant difference according to the G-test). Interestingly, these two events are by far the most likely to have titles or subtitles (i.e., where self-description is included, it is often included as text in addition to or instead of narration) and also by far the most likely to be music-tracked (see Table~\ref{tab:postprodpos})---though apparently not necessarily music with lyrics .

Table~\ref{tab:spokenbyevent} also breaks down the types of languages spoken or sung for each event. There is significant variation between events in how dominant English-language videos are, ranging from 81\% of positive workworking-project videos with spoken language to 97\% of positive animal-feeding videos (p $<$ 0.0001 for English (only) vs. identified and non-identified non-English (only), chi-square). It is not clear to what effect the distribution of languages across events in YLI-MED is an effect of how the dataset was collected, as opposed to an effect of which language speakers upload each type of video.
For example, the variation might in part be an effect of how difficult it was to find videos for that event (as discussed in \S\ref{sec:collection}); for the easier events, multilingual searching was an add-on rather than a necessity (though many non-English videos were gathered with English search terms anyway). For the easier events (listed in Table~\ref{tab:excludedterms}), the percentage of videos with English (or English and some other language) is slightly higher, at 92\% on average out of positive videos with spoken language, as opposed to 88\% on average for the more difficult-to-find videos. However, this effect is not statistically significant (p = 0.5959, Fisher's exact test).

\begin{table*}[htp]
\centering
{\renewcommand{\arraystretch}{1.2}
\begin{tabular}{| p{0.35in} | p{0.95in} || p{0.35in} | p{0.35in} || p{0.575in} | p{0.575in} | p{0.575in} | p{0.575in} | p{0.575in} | }
\hline
Event & Event & \multicolumn{2}{c||}{Spoken Lang?} & \multicolumn{5}{c|}{Within Videos That Have Spoken/Sung Language} \\ \cline{3-9}
\# & Name & Yes & No & English & Other Identified Language & Non-English (Unidentified) & Multiple (English \& Other) & Unintel-ligible \\ \hline
Ev101 & Birthday Party & 98.7\% & 1.3\% & 91.5\% & 5.1\% & 1.7\% & 1.7\% & 0.0\% \\ \hline
Ev102 & Flash Mob & 99.3\% & 0.7\% & 89.9\% & 0.0\% & 10.1\% & 0.0\% & 0.0\% \\ \hline
Ev103 & Vehicle Unstck & 99.3\% & 0.7\% & 81.4\% & 2.9\% & 14.3\% & 1.4\% & 0.0\% \\ \hline
Ev104 & Parade & 96.5\% & 3.5\% & 89.2\% & 5.0\% & 4.1\% & 0.0\% & 1.8\% \\ \hline
Ev105 & Board Trick  & 90.2\% & 9.8\% & 87.4\% & 2.3\% & 7.4\% & 2.9\% & 0.0\% \\ \hline
Ev106 & Groom Animal & 97.8\% & 2.2\% & 95.6\% & 2.2\% & 2.2\% & 0.0\% & 0.0\% \\ \hline
Ev107 & Feeding Animal & 99.1\% & 0.9\% & 96.8\% & 0.0\% & 2.8\% & 0.5\% & 0.0\% \\ \hline
Ev108 & Landing a Fish & 99.3\% & 0.7\% & 90.8\% & 0.7\% & 8.5\% & 0.0\% & 0.0\% \\ \hline
Ev109 & Wedding & 100.0\% & 0.0\% & 88.1\% & 0.0\% & 11.9\% & 0.0\% & 0.0\% \\ \hline
Ev110 & Woodworking & 89.3\% & 10.7\% & 80.6\% & 6.7\% & 11.9\% & 0.7\% & 0.0\% \\ \hline
\multicolumn{2}{|l||}{Total: Positives} & 97.0\% & 3.0\% & 89.5\% & 2.5\% & 7.0\% & 0.7\% & 0.2\% \\ \hline
\multicolumn{2}{|l||}{Total: All Reviewed} & 95.9\% & 4.2\% & 88.7\% & 3.2\% & 7.0\% & 0.7\% & 0.4\% \\ \hline
 \end{tabular}}
\caption{Percentage of positive-example videos that have spoken or sung language by event, and breakdown by language type within those that have spoken/sung language.}
\label{tab:spokenbyevent}
\end{table*}

The language of any text added in post-production (titles, subtitles, etc.) seems to show a similar patterning to spoken language across the different target events, at least for those events where there are more than a few videos with added text at all. However, since the overall numbers of videos with added text are relatively small, it is difficult to assess whether any apparent differences between events are significant.  (If it were, again, it would probably reflect more about the collection process than about the ``natural'' distribution of the data.) 

\subsection{Comparing the Training and Test Sets}
\label{sec:ttcomp}

This section compares the standard Training and Test sets of positive event videos in the YLI-MED corpus to determine whether there are any statistically significant differences between them in terms of corpus-builders' category judgments or other annotations.\fn{Negative videos are not included here because they were not annotated for any of the relevant characteristics. As we noted in \S\ref{sec:nonpos}, the Supplemental set of near misses and related videos is small and was not specifically composed to be balanced in any way, so corpus users should not expect it to be statistically comparable to the main Training and Test sets.} No attempts have been made to compensate for any of the differences found, which are generally quite minor even where statistically significant. (Determining whether such differences affect measurements of classifier performance may be an interesting topic for research.)

Table~\ref{tab:compagrconf} shows the degree of annotator agreement on whether a video was a positive example (2/3 vs.\ 3/3) and the average confidence among annotators judging it as positive for the Training and the Test sets as a whole (all events). It also shows that the differences in agreement are not statistically significant, nor are the differences in average confidence.\fn{Except as otherwise noted, Fisher's exact test, two-tailed, was used to assess the statistical significance of variation in annotator agreement on relation (i.e., on whether a video was a positive instance), and Student's unpaired t-test, two-tailed, was used to assess the significance of variation in annotator confidence.} 

\begin{table*}[htp]
\centering
{\renewcommand{\arraystretch}{1.2}
\begin{tabular}{| p{1.25in} | p{0.825in} | p{0.825in} | p{0.825in} | p{0.825in} | p{0.825in} | }
\hline
Set (Positives) & \% Partial Agreement & Avg. Confidence for Partial & \% Full Agreement & Avg. Confidence for Full & Total Avg. Confidence \\ \hline
Training (all events) & 7.7\% & 1.97 & 92.3\% & 2.77 & 2.71 \\ \hline
Test (all events) & 6.8\% & 1.94 & 93.2\% & 2.77 & 2.71\\ \hline
Significance & p=0.4710 & p=0.8020 & p=0.4710 & p=0.8939 & p=0.9292 \\ \hline
\end{tabular}}
\caption{Comparison of annotator agreement and confidence in positive judgments for Training vs.\ Test sets.}
\label{tab:compagrconf}
\end{table*}

Testing the differences between the Training and Test sets for individual events, we found no significant differences in level of agreement for any event except Ev107 {\it Person Hand-Feeding an Animal}, which shows a marginally significant difference (p = 0.0237). There are also no significant differences in total average confidence between the Training and Test sets for any individual events except Ev105 {\it Attempting a Board Trick} (p = 0.0102) and Ev108 {\it Person Landing a Fish} (p = 0.0010).\fn{We did not check the statistical significance of difference in confidence within each agreement category (partial vs.\ full) for the individual events. However, it is worth noting that, for each of the events, the differences in average confidence between the Training and Test sets within an agreement category are always smaller than the difference in confidence {\it between} partial and full agreement. Average confidences for Training and Test are generally closer within the full-agreement categories, which is not surprising given how many more videos they comprise.} 

Table~\ref{tab:comppostprod} shows the percentages of the Training, Test, and Supplemental sets that have certain production or post-production characteristics. Comparison of the positive Training and Test sets shows that differences in the proportion of animated videos (2\% vs.\ 1\%) and videos with musical scores (3\% vs.\ 4\%) are not statistically significant (Fisher's exact test, two-tail). Differences in the proportion of videos with added text (5\% vs.\ 3\%) are only marginally significant. Within those videos with an added music track, there is no statistically significant difference between the Training and Test sets in whether the music track replaces the original audio, is overlaid on top of it, or is interspersed with it (p = 0.3041, Freeman-Halton extension of Fisher's exact).

\begin{table*}[htp]
\centering
{\renewcommand{\arraystretch}{1.2}
\begin{tabular}{ | p{1.4in} | p{1.4in} | p{1.4in} | p{1.4in} | }
\hline
Set (Positives) & Animation/CGI & Added Text & Musical Score \\ \hline\hline
Training & 1.7\% & \textbf{4.7\%} & 3.0\% \\ \hline
Test & 1.0\% & \textbf{2.9\%} & 4.3\% \\ \hline
Significance & p=0.2262 & \textbf{p=0.0522} & p=0.1637 \\ \hline\hline
Supplemental Set & 0.6\% & 8.5\% & 6.8\% \\ \hline
\end{tabular}}
\caption{Comparison of post-production characteristics for positive videos in the Training vs.\ Test sets, vs.\ the Supplemental set of near misses and related videos. (Marginally significant differences are shown in boldface.)}
\label{tab:comppostprod}
\end{table*}

Table~\ref{tab:traintestlg} compares the percentages of the Training and Test sets that have spoken or sung language and the type of language they have. There is a slight, marginally significant difference in how many videos have spoken or sung language, with 98\% in the Training set and 96\% in the Test set (p = 0.0376, Fisher's exact). Among those videos that have spoken/sung language, the distribution of languages is very similar, with no significant differences (p = 0.8704 for English (only) vs. identified and non-identified non-English (only), Fisher's exact; p = 0.2046 across all categories, chi-square).

\begin{table*}[htp]
\centering
{\renewcommand{\arraystretch}{1.2}
\begin{tabular}{| p{0.55in} || p{0.5in} | p{0.5in} || p{0.7in} | p{0.7in} | p{0.7in} | p{0.7in} | p{0.7in} | }
\hline
Set & \multicolumn{2}{c||}{Spoken Language?} & \multicolumn{5}{c|}{Within Videos That Have Spoken/Sung Language} \\ \cline{2-8}
(Positvs) & Yes & No & English & Other Identified Language & Non-English (Unidentified) & Multiple (English \& Other) & Unintel-ligible \\ \hline
Training & \textbf{97.8\%} & \textbf{2.2\%} & 90.1\% & 2.1\% & 7.3\% & 0.4\% & 0.1\% \\ \hline
Test & \textbf{96.1\%} & \textbf{3.9\%} & 88.9\% & 2.9\% & 6.7\% & 1.1\% & 0.4\% \\ \hline
 \end{tabular}}
\caption{Comparison of presence of spoken or sung language in positive videos in the Training vs.\ Test sets, and breakdown by language type within those that have spoken/sung language. (Marginally significant differences in boldface.)}
\label{tab:traintestlg}
\end{table*}

{\it User Distribution:} As we noted in \S\ref{sec:splitting}, the distribution of videos per user was manipulated to allow more unevenness in the training set than the test set; a comparison is given above in Table~\ref{tab:usersplit}.

\subsection{A Mini-Analysis: Correlations Between Non-Event Characteristics and Event Decisions}
\label{sec:psych}

In multimedia research, it is well-known that certain production and post-production characteristics can have a significant effect on performance for automated video content analysis of various kinds. It is therefore of interest to investigate whether these particular factors also affect how easy it is for humans to classify videos, in our case, by the events occurring in them. 

While our video-collection process does not constitute a controlled psychological experiment, it does offer some intriguing data on how humans interpret video that may be of specific interest to multimedia researchers. This subsection therefore offers a few preliminary findings based on the corpus-builders' agreement on whether videos were positive exemplars of their target events or whether they were near misses or simply related, and on their confidence in those judgments.\fn{All statistics presented in this section are for videos on which the consensus among annotators is that it is a Positive instance.} 

These results are particularly interesting for researchers exploring how to use audio vs.\ visual information in multimedia event detection, and how to combine information from each domain. 

\subsubsection{Effects of Post-Production}
\label{sec:postprodFX}

Table \ref{tab:postprodeffects} shows the effects of selected post-production characteristics on agreement between YLI-MED annotators about how to categorize videos and on their confidence in those judgments.\fn{For all of the statistics presented in this subsection, Fisher's exact test, two-tailed, was used to assess the statistical significance of variation in annotator agreement on relation, and Student's unpaired t-test, two-tailed, was used to assess the significance of variation in annotator confidence.} 

\begin{table*}[htp]
\centering
{\renewcommand{\arraystretch}{1.2}
\begin{tabular}{| p{1.5in} || p{0.55in} | p{0.55in} | p{0.75in} || p{0.5in} | p{0.5in} | p{0.75in} | }
\hline
\multirow{2}{*}{}  & \multicolumn{3}{c||}{\% Full Agreement} & \multicolumn{3}{|c|}{Average Confidence} \\ \cline{2-7}
 & Y & N & Significance & Y & N & Significance \\ \hline
Animation & 84.0\% & 92.8\% & p=0.1037 & 2.63 & 2.71 & p=0.4856 \\ \hline
Added Text & 88.7\% & 92.9\% & p=0.2376 & 2.72 & 2.71 & p=0.8056 \\ \hline
Musical Score & \textbf{\textit{83.1\%}} & \textbf{\textit{93.1\%}} & \textbf{\textit{p=0.0061}} & \textbf{2.56} & \textbf{2.71} & \textbf{p=0.0204} \\ \hline
\ \ \ Overlaid & 88.6 & 93.1\% & p=0.5271 & 2.62 & 2.71 & p=0.3221 \\ \hline
\ \ \ Interspersed & 100.0 & 93.1\% & p=1.0000 & 3.00 & 2.71 & p=0.1851 \\ \hline
\ \ \ Score Only & \textbf{\textit{70.8\%}} & \textbf{\textit{93.1\%}} & \textbf{\textit{p=0.0011}} & \textbf{\textit{2.35}} & \textbf{\textit{2.71}} & \textbf{\textit{p=0.0009}} \\ \hline
\end{tabular}}
\caption{Effects of post-production characteristics on annotator agreement and average confidence per video, for all positive-example videos in the YLI-MED corpus. (Significant effects are shown in boldface italics, with marginally significant effects in boldface alone.)}
\label{tab:postprodeffects}
\end{table*}

Animation did not have a statistically significant effect on either agreement about whether a video is a positive example (full 3/3 vs.\ partial 2/3) nor on average annotator confidence (among those judging it positive).

Taken as a whole, text added in post-production did not have a significant effect on either agreement or annotator confidence. However, closer examination shows that the effects may depend on what the language is; this is discussed in detail in \S\ref{sec:lgeffects}.

Musical scoring, on the other hand, had {\it a significant effect} on agreement, with 83\% of scored videos achieving full agreement as opposed to 93\% for unscored videos. Musical scoring also had a marginally significant effect on annotator confidence, at an average of 2.56 for scored videos and 2.71 for unscored. However, further investigation reveals that the significant effect stems solely from those cases where the music track entirely replaces any original audio; the latter case has {\it highly significant effects} on both agreement and confidence. The the other two music-score types we annotated (overlaid music and interspersed music) did not produce any statistically significantly different effect on agreement or confidence when compared to non-music-tracked videos.\fn{Table~\ref{tab:postprodeffects} compares subtypes of musical scoring with videos having no score, but similar significance effects are obtained by comparing each subtype to all other videos; the videos where any original audio track has been completely replaced are producing the effect. However, differences in agreement {\it between} the three subtypes are not statistically significant ($P_A$ and $P_B$=0.1479, Freeman-Halton extension of Fisher's exact test).}

\subsubsection{Effects of Language}
\label{sec:lgeffects}

Table \ref{tab:spokenlgeffects} breaks down degree of annotator agreement (full 3/3 as opposed to partial 2/3) and average confidence scores for the positive examples by whether the video's audio track contained spoken (or sung) language and what language it was.\fn{As described in \S\ref{sec:language}, the label `Unintelligible' is used where the annotators could not make out the language well enough to tell whether it was English, while `Non-English' is used where annotators were sure the language wasn't English, but didn't know what it was.} However, not all of the apparent effects are statistically significant.

\begin{table*}[htp]
\centering
{\renewcommand{\arraystretch}{1.2}
\begin{tabular}{| p{3in} | p{0.7in} | p{0.9in} | p{0.9in} |}
\hline
 & N & \% Full Agreement & Average Confidence \\ \hline
No Spoken Language & 54 & 85.2\% & 2.49 \\ \hline
Spoken Language & 1769 & 92.9\% & 2.71 \\ \hline
\ \ \ Intelligible ($\geq 1$ Language) & \ \ \ 1765 & \ \ \ 92.9\% & \ \ \ 2.71 \\ \hline
\ \ \ \ \ \ Specifically Identified ($\geq 1$ Language) & \ \ \ \ \ \ 1641 & \ \ \ \ \ \ 93.3\% & \ \ \ \ \ \ 2.74 \\ \hline
\ \ \ \ \ \ \ \ \ English (Only) & \ \ \ \ \ \ \ \ \ 1584 & \ \ \ \ \ \ \ \ \ 93.4\% & \ \ \ \ \ \ \ \ \ 2.74 \\ \hline
\ \ \ \ \ \ \ \ \ Other Identified Language & \ \ \ \ \ \ \ \ \ 44 & \ \ \ \ \ \ \ \ \ 90.9\% & \ \ \ \ \ \ \ \ \ 2.73 \\ \hline
\ \ \ \ \ \ \ \ \ Multiple Languages & \ \ \ \ \ \ \ \ \ 13 & \ \ \ \ \ \ \ \ \ 84.6\% & \ \ \ \ \ \ \ \ \ 2.64 \\ \hline
\ \ \ \ \ \ Non-Identified (Non-English) & \ \ \ \ \ \ 124 & \ \ \ \ \ \ 87.9\% & \ \ \ \ \ \ 2.43 \\ \hline
\ \ \ Unintelligible & \ \ \ 4 & \ \ \ 100.0\% & \ \ \ 2.92 \\ \hline
\end{tabular}}
\caption{Breakdown of the effects of spoken (or sung) language on annotator agreement and average confidence per video, for all positive-example videos in the YLI-MED corpus.}
\label{tab:spokenlgeffects}
\end{table*}

The presence of spoken or sung language in the audio content of the video has a marginally significant effect on agreement, with 93\% full agreement on videos with audible language vs.\ 85\% full agreement on videos without audible language (p = 0.0544). More notably, it has {\it a highly significant effect} on annotator confidence, with an average of 2.71 (out of 3.00) annotator confidence vs.\ 2.49 (p = 0.0022). Further comparison of the effects within those videos with audible language shows that:

\bi
\item Among those videos where the language was intelligible, the most significant effects arose from whether or not the language was identifiable (whatever the identified language was). Identifiability had a marginally significant effect on annotator agreement, at 93\% full agreement for identified languages and 88\% agreement for languages none of the corpus builders could identify (p = 0.0406). It had {\it a highly significant effect} on annotator confidence, at 2.74 vs. 2.43 (p $<$ 0.0001). This may be because the corpus-builders understood the languages they could identify (even if not English), and thus had more evidence to support event categorizations, or it may simply be that being able to identify the language in the video generally gave them a more confident feeling.\fn{It is worth keeping in mind that the corpus-builders do not all speak/understand/recognize the same set of languages, so it might be the case that, even for a video with a specific identified language annotation, only one of the annotators recognized the language.}
\item However, comparing only identified non-English languages with unindentified non-English languages, the effect is not as strong, statistically speaking; there is {\it a significant effect} on confidence (p = 0.0105) but not on agreement (p = 0.7833).
\item On the other hand, it is not the case that the identifiability effect is all due to English; comparing English with identified non-English videos, we find that there is no significant difference between them (p = 0.5307 for agreement and p = 0.9503 for confidence). This suggests a subtle interaction between general identifiability and Englishness in particular.
\item For videos with multiple spoken or sung languages, one of the languages was usually English, and indeed, multiple-language videos pattern statistically with English videos in significance tests.
\item The effect of intelligibility is unclear; given the low number of truly unintelligible videos, comparisons were not statistically significant.
\ei

As noted in \S\ref{sec:postprodFX}, visual text added to the video in post-production did not in and of itself have a statistically significant effect on annotator agreement nor confidence. However, a closer look reveals that, within the videos that do have added text, there are some differences depending on what language the text is in---though, because the number of videos with added text is relatively few, it is more difficult to identify significant trends than with the spoken-language data. Table~\ref{tab:textlgeffects} breaks down the numbers given in \S\ref{sec:postprodFX} into language categories.

\begin{table*}[htp]
\centering
{\renewcommand{\arraystretch}{1.2}
\begin{tabular}{| p{3in} | p{0.7in} | p{0.9in} | p{0.9in} |}
\hline
 & N & \% Full Agreement & Average Confidence \\ \hline
No Added Text & 1748 & 92.9\% & 2.71 \\ \hline
Added Text & 75 & 88.0\% & 2.65 \\ \hline
\ \ \ Actually Language ($\geq 1$ Language) & \ \ \ 74 & \ \ \ 87.8\% & \ \ \ 2.65 \\ \hline
\ \ \ \ \ \ Specifically Identified ($\geq 1$ Language) & \ \ \ \ \ \ 73 & \ \ \ \ \ \ 89.0\% & \ \ \ \ \ \ 2.64 \\ \hline
\ \ \ \ \ \ \ \ \ English (Only) & \ \ \ \ \ \ \ \ \ 66 & \ \ \ \ \ \ \ \ \ 90.9\% & \ \ \ \ \ \ \ \ \ 2.64 \\ \hline
\ \ \ \ \ \ \ \ \ Other Identified Language & \ \ \ \ \ \ \ \ \ 6 & \ \ \ \ \ \ \ \ \ 66.7\% & \ \ \ \ \ \ \ \ \ 2.67 \\ \hline
\ \ \ \ \ \ \ \ \ Multiple Languages & \ \ \ \ \ \ \ \ \ 1 & \ \ \ \ \ \ \ \ \ 100.0\% & \ \ \ \ \ \ \ \ \ 3.00 \\ \hline
\ \ \ \ \ \ Non-Identified (Non-English) & \ \ \ \ \ \ 1 & \ \ \ \ \ \ 0.0\% & \ \ \ \ \ \ 3.00 \\ \hline
\ \ \ Proper Names Only & \ \ \ 1 & \ \ \ 100.0\% & \ \ \ 3.00 \\ \hline
\end{tabular}}
\caption{Breakdown of effects of the language of post-production text on annotator agreement and average confidence per video, for all positive-example videos in the YLI-MED corpus.}
\label{tab:textlgeffects}
\end{table*}

While spoken language generally had more significant effects on confidence than on agreement, the language of post-production text had some effect on agreement but none at all on confidence. Specifically:

\bi
\item Videos with English text, which had a full-agreement rate of 93\%, showed a marginally significant difference from those with non-English text (57\% full agreement, combined), whether the annotators could identify the language of the text or not (p = 0.0361). However, no significant effects were found for either case alone (identified vs.\ unidentified non-English text), likely because the numbers are so small.
\item Differences between videos with no added text and videos with added English text were not statistically significant (p = 0.4684 for agreement).  However, taken together, videos with no added text and videos with English text showed {\it a significant difference} in annotator agreement compared with non-English-text videos (p = 0.0107). This patterning is likely because most of the videos with no added text had spoken English in any case (as that describes the vast majority of the positives in the corpus).
\ei

\section{Limitations, Corpus Comparability, and Generalizability}
\label{sec:comparability}

Any set of procedures involves choices that will result in biases in the product. Our goal in this report is to document those choices as best we can, so that corpus users will be aware of any potential biases. In \S\ref{sec:biasesdisc}, we discuss some possible biases in detail, along with potential consequences for the generalizability of classifiers trained on YLI-MED in comparison with other datasets. In particular, we note some potential differences that may have been introduced between YLI-MED and the Heterogenous Audio Visual Internet Collection (HAVIC) (\citealt{HAVIC}), which YLI-MED intentionally parallels in many respects, as a result of the differences between our procedures and those of the LDC that we noted in \S\ref{sec:procedures}. In \S\ref{sec:havicdiffs}, we point out some differences between YLI-MED and HAVIC that are intentional, not mere side effects.

Quantifying potential distributional biases in YLI-MED in comparison with other corpora---and their potential effects on classifiers trained therewith---is largely an area for future research. However, this section offers some hints that may nonetheless be helpful for corpus users, and in \S\ref{sec:compclassifiers}, we make some preliminary comparisons of results for event detectors trained and tested on YLI-MED vs.\ TRECVID MED data from HAVIC.

\subsection{Potential Unintentional Biases}
\label{sec:biasesdisc}

\subsubsection{Biases Arising from the YFCC100M Dataset}

Some of the potential differences between YLI-MED and similar corpora arise from our use of the Yahoo Flickr Creative Commons 100 Million dataset as a basis from which to select videos. While the YFCC100M's nearly 800,000 videos constitute the largest existing open-source research dataset of consumer-produced videos, and therefore contains quite a large variety of video types and topics, it is still a product of a particular community of users with a particular geographic, cultural, and linguistic distribution, and represents those users' interests and styles. In fact, even at the scale of YFCC100M, there can be skew towards particular prolific users; as we noted in \S\ref{sec:peruser}, 5.8\% of the videos in YFCC100M were contributed by just 10 users.

A comparison with YouTube videos from a similar time period shows that the distribution of tags is fairly comparable, though YouTube videos had an average of twice as many tags per video (\citealt{YFCC2014}). There are a few idiosyncrasies in tagging behavior on Flickr, such as using tags for dates and community awards (for both images and videos). However, for the most part, the tag distribution for these two U.S.-based video-sharing services is similar enough to indicate that the common topics are largely shared (at least among tagged videos). Of course, it is an open question how similar either would be to, say, a China-based provider like Tudou or Youku---and what effect cultural specificity might have on the models developed and thus the potential cross-cultural/cross-dataset performance of classifiers for each event.

Beyond biases inherent to Flickr, there may be biases in the particular subset  used for the YFCC100M. In particular, Creative Commons--licensed videos are produced by a particular community of users, who may, for example, be less likely to be professional photographers or videographers than users who maintain full copyright. 
In addition, YFCC100M is restricted to videos meeting Flickr's ``Safe Search'' criteria and only covers videos uploaded through March 2014, though it is unclear what effect either of these restrictions might have on representativity with respect to our target events. Most UGC video corpora have similar restrictions, i.e., excluding copyright-restricted and adult-content videos, and of course each was gathered over a particular time period.

\subsubsection{Possible Effects of Procedure on Representativity: YLI-MED vs.\ YFCC100M}
\label{sec:procchoices1}

The most obvious overall bias in YLI-MED is that the reviewed videos are more likely to have English-language tags than the YFCC100M as a whole. As we described in \S\ref{sec:collection}, the video collectors were native English speakers and for the most part relied on English search terms---and efforts to search in other languages had less chance of success, due both to the searchers' not knowing those languages (or not knowing them well) and to the difficulty of searching in highly inflected languages. However, many videos with non-English content are tagged in English---in fact, English tags dominate by orders of magnitude on Flickr (at least as of \citealt{folks}).\fn{Although we have not done a thorough analysis for YFCC100M like that of \citeauthor{folks}, a glance at the top 100 tags reveals that all the common nouns that are uniquely categorizable as to language are English.}). Therefore, we assume that tag-induced skew toward English-language content is not as great as it might otherwise be (see \S\ref{sec:language} for a breakdown of content-language distribution for the YLI-MED positives).

\begin{figure}[thp]
\centering
\includegraphics[width=6.5in]{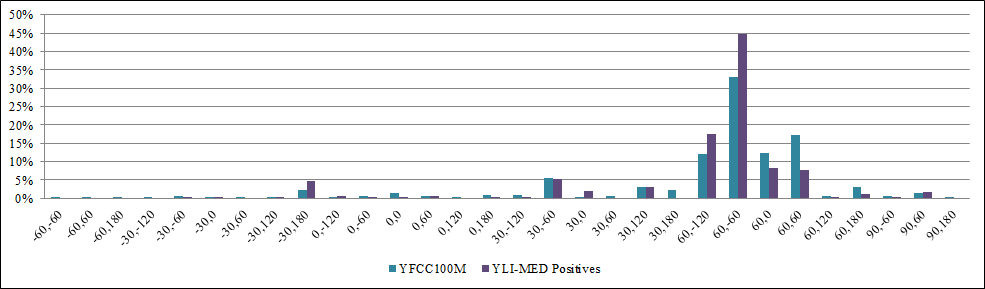}
\caption{Comparison of the proportional geographical distributions of geotagged positive-example videos in YLI-MED vs.\ all videos in YFCC100M, in terms of grid placement. X-axis labels identify the latitude and longitude of the northeast corner of each grid unit. }
\label{fig:geobreakdown}
\end{figure}

We cannot precisely quantify language skew without knowing the language distribution of the full YFCC100M; however, geography may provide a proxy. Figure~\ref{fig:geobreakdown} compares the distribution of the 476 geotagged videos among the positive examples in YLI-MED with the distribution of the 103,505 geotagged videos in the whole YFCC100M, using a very rough division of the globe into 30-degree-by-60-degree grid units.\fn{The YLI-MED positive videos are twice as likely to be geotagged as YFCC100M videos in general, at 26\% to 13\%. This is not surprising, as the reviewed videos in YLI-MED must generally be likely to have more user-supplied metadata.} This first-pass comparison shows that the proportion of videos from the (predominantly English-speaking) U.S. and southern Canada (60,-60 in the figure) is indeed notably higher among YLI-MED positives than in YFCC100M as a whole, as is the proportion of videos from Australia (-30,180). The proportion of videos from (multilingual) Europe, the Middle East, and northern Africa (60,0 and 60,60 in the figure) is notably lower among YLI-MED positives than for YFCC100M as a whole.

Among the non-English videos, it is possible that there is more of a bias towards those that contain more prototypical examples of events (or more prototypical to an American), or where the nonverbal evidence is clearer, in comparison with videos in English. In videos in English (or other languages they knew), the corpus builders could use evidence from what was said in the video (or the metadata, in borderline cases) to determine that an event was occurring, even if the nonverbal evidence was not clear or the activities were not prototypical for the event. 

For some events---especially those listed in Table~\ref{tab:excludedterms}, where the YFCC100M contained many more relevant videos than were needed to make up the YLI-MED event set---the distribution may specifically be skewed towards videos tagged with the most obvious search terms or with the (English) search terms that the collectors happened to try first (often a word or words from the event name, for example {\it wedding} for Ev109 {\it Wedding Ceremony}). While collectors were encouraged to try a variety of search terms in a variety of languages, this did not always work out in practice---if a search term has returned numerous relevant results, it seems more efficient to continue reviewing those results than to try a different term that might be less productive.
Depending on how the first terms tried relate to the overall distribution of terms in the metadata, this happenstance may (or may not) result in models that overfit to some particular subset of the videos for that event (i.e., the subset that would be tagged with those terms).\fn{Again, we have not yet conducted experiments to determine whether these possible biases have any effect on classifier performance; we are simply noting some potential issues corpus users may wish to keep in mind.} 

Skew towards the first search terms tried is less likely to be a problem for those events that were more difficult to find videos for, as collectors used a greater variety of terms, in more languages (thus likely getting a more representative distribution from the YFCC100M). On the other hand, the videos for more difficult-to-find events are more likely to be skewed towards particular users, as we were not able to reduce the number of videos per user as easily where we did not have as many videos to start with. See \S\ref{sec:userbias} for a detailed discussion of skew towards prolific uploaders as another potential source of corpus bias, and \S\ref{sec:extratrick} for an invitation to explore and better quantify the potential effects.

\begin{figure}[thp]
\centering
\includegraphics[width=6.5in]{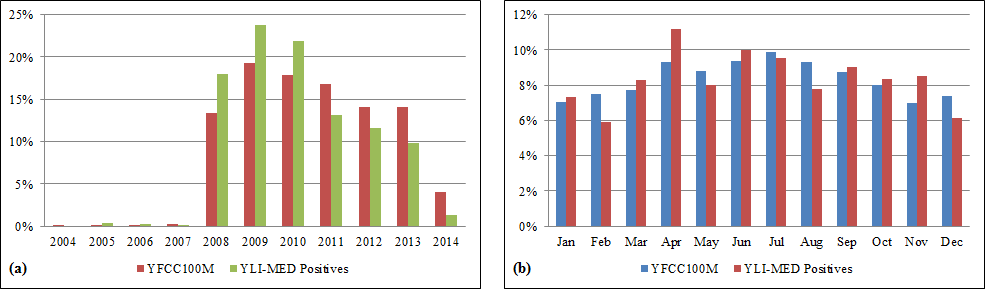}
\caption{Comparison of the proportional temporal distributions of positive-example videos in YLI-MED vs.\ all videos in YFCC100M, (a) by upload year and (b) by upload month (for all years combined).}
\label{fig:uploadtime}
\end{figure}

Again, we have not comprehensively studied how these potential biases play out in the composition of the YLI-MED corpus, other than the few comparisons with the numerical makeup of the YFCC100M we have mentioned. Another such quantitative comparison can be made for distribution over time. Figure~\ref{fig:uploadtime} shows the distribution of videos across years and across months throughout the year.\fn{The data in Figure~\ref{fig:uploadtime} are for YLI-MED Version 1.1. Videos with errors in the upload timestamp are excluded.} For both time dimensions, the positive-example videos in YLI-MED show similar general patterns to the YFCC100M as a whole, but there is more variability for YLI-MED (and preliminary significance testing shows those differences are statistically significant).

However, it is not clear what the source of the discrepancy is; in fact, it seems {\it a priori} unlikely that upload-time distribution should be affected by corpus collection procedures (though there may be factors we haven't thought of). But temporal distribution may well be affected by the nature of the target events themselves, in that the relevant activities may be seasonal or may change in popularity over time---or the greater variability of YLI-MED may simply be a random effect, due to the small size of the subset.

\subsubsection{Possible Effects of Procedure on Comparability: YLI-MED vs.\ TRECVID MED}
\label{sec:procchoices2}

The potential search-term skew described above arises in part from the fact that, as described in \S\ref{sec:collection}, the bulk of the initial video collecting was performed by one person. This is, in fact, one of the largest differences between our procedures and those used to collect the HAVIC corpus: while we were optimizing for efficiency (in time and money), HAVIC was a much larger-scale operation, with the work distributed among many data scouts. Among other things, this distributed structure had the advantage that the LDC team could add mechanisms such as games and contests that explicitly rewarded variety (\citealt{HAVIC}). 

In addition, as we noted in \S\ref{sec:definitions}, our small scale dictated that our event definitions (and the research that went into them) focus more on what to do with the edge cases (i.e., on how to decide what ``counts'' as an example of a given event), to discourage corpus-builders from applying narrow criteria that did not accurately reflect the breadth of other Internet users' conceptual models of that event. (Whereas, while a given HAVIC data scout's model of an event might deviate significantly from that of most other Internet users', having several scouts for each event is likely to have largely mitigated skew towards idiosyncratic models.) This explicit direction on edge cases, combined with the fact that each candidate video was viewed by three people rather than HAVIC's two, may have resulted in more homogeneity per event in YLI-MED. On the other hand, as we did include videos where only two of the three annotators agreed, the effect may be small, nonexistent, or even the opposite.

In creating the event definitions, we aimed to reflect the sense of the event names intended by the LDC (modulo allowing for non-human protagonists). However, because the HAVIC event definitions are proprietary, we were unable to simply adapt those definitions, nor directly compare ours with them. It may be the case that small differences in the content of the core definitions---perhaps more importantly than their focus---led to the YLI-MED collectors choosing different videos than they would have if they were working with the HAVIC definitions.\fn{As we describe in Appendix~\ref{sec:defsapp}, the definition for Ev107 {\it Person Hand-Feeding an Animal} unintentionally implied that tossing food to a fish or bird would never count (contra HAVIC), and some of the corpus-builders did interpret it that way. However, the event-detection data discussed in \S\ref{sec:compclassifiers} show that the substantively different definitions for that event resulted in sets of videos that an i-vector-based detection system modeled comparably well (more comparable than for most of the events).} 
It is also of course quite possible that, despite the abovementioned guidance on edge cases, the YLI-MED video sets for some events may be skewed towards corpus-builders' individual models of the events---or that the given definitions reflect some skew in the data and methods used to construct them. 

Finally, our procedures for developing the ``negatives'' set (Ev100) were considerably different than those used for HAVIC, in that we did not individually review each of the negative videos to ensure that it really was not an event video.\fn{However, \citeauthor{HAVIC} point out that, due to ongoing changes in the process and target events, HAVIC's negative set is not free of accidental positives either (\citeyear{HAVIC}).} (Again, due to practicalities of scale.)
While we did take other steps to try to screen out possible event videos, as described in \S\ref{sec:negatives}, it is likely that up to 1.6\% of the videos labeled {\it None of the Above} are actually positive instances, near misses, or related videos---especially for the events most heavily represented in the YFCC100M (see Table~\ref{tab:excludedterms}). These miscategorized videos will share the characteristic of not being easy for an English speaker to search for based on textual metadata  (see \S\ref{sec:unavoidables}).\fn{In addition to the known procedural differences discussed in this section, there are some potentially biasing procedural points where we don't know whether YLI-MED and HAVIC differed. For example, it is not clear from the published documentation whether HAVIC data scouts were specifically encouraged to use non-English search terms, as ours were (though HAVIC certainly includes non-English videos), nor is it clear whether there were limits on videos per user for the TRECVID MED data.} 

One area in which both YLI-MED and HAVIC may differ from other corpora is that the near misses and related videos are separated from the true negatives (which are completely unrelated to any target event). This necessarily means that the negatives set contains fewer of these borderline cases, and that researchers need to decide whether to explicitly include them in a given training or test run. However, given the relatively small number of near misses and related videos included in YLI-MED, and the fact that the negatives have not been exhaustively cleared of such cases, we assume that there are still many near misses and related videos in the negatives set (especially for the videos where we did not have to cast as wide a search net to find the positives). 
In addition, the fact that the negative sets for both corpora (mostly) do not include videos with the target events may skew the overall subject matter of the negatives somewhat away from the domains in which those events are likely to occur. For example, NIST's (and thus our) events are less likely to involve the commercial, military, work, or entertainment domains. 

\subsubsection{Are There Biases No Search Method Can Avoid?}
\label{sec:unavoidables}

There are some ways in which any event corpus constructed by searching on the user-supplied metadata for user-generated content would likely be skewed, i.e., would not reflect the actual distribution of videos depicting those events along all possible dimensions throughout the source data or in comparison with all of the online videos in existence.\fn{The least bias-inducing method, of course, would be to simply review every video in the dataset (or every video on the web) to see whether it depicted any of the target events. This is not feasible to do every time one wants to develop a new set of event classifiers. However, there are some reasonable compromises: blanket annotation/tag refinement of a corpus with a rough idea of what it might be used for can at least provide better searchability than relying solely on user-supplied metadata. We are therefore moving towards providing a full set of basic annotations for the YFCC100M: essentially, a Multimedia Genome Project (see \S\ref{sec:nextsteps}).} These factors will not necessarily make YLI-MED different from other UGC video corpora (all constructed from such searches), but some are nonetheless worth noting.

To begin with, such a corpus is not going to include videos where the user simply did not supply metadata. In addition, it will include (mostly) videos where at least some of the textual metadata is descriptive of the content of the video---rather than, for example, descriptions of the camera, reasons for posting (``Thought this was interesting''), or random unrelated rants (\citealt{ulges2010visual}). Further research is necessary to discover what the consequences of these factors may be for corpus composition and distribution, but possibilities that might be explored include biases against politically subversive content (which a poster might not want to make directly searchable), against highly artistic videos (where the aesthetic might be more noteworthy than what is depicted), or against didactic videos (where the metadata might reflect the content taught rather than what occurs in the video). 

Even where the textual metadata is descriptive of the video content, the video will not be found if it doesn't match the aspect of the video the searcher is looking for or is not phrased in a way the searcher might expect. For example, the description ``Me and Sasha at Dynamo's last Friday'' is quite specific, but will not be helpful to someone searching for event examples. Previous research has shown that Flickr users are more likely to tag people and places than story content in videos (\citealt{flickrvindexing}), indicating that events may generally be one of the less searchable aspects of videos in the first place.

One very speculative possibility is that an event corpus might contain fewer videos that feature particularly notable places and people, if notability affects which tags are chosen (and what shows up in titles, etc.), at least by less voluble uploaders. (If you get a shot of Beyonc\'{e} digging her car out of a snowbank, is your description more likely to feature the stuck vehicle or Beyonc\'{e}? Would it be different for a shot of your neighbor doing the same thing?) Again, these are all empirical questions, which we hope to explore in future research.

In addition, videos found through textual metadata may be less likely to be from batch uploads. For example, if someone is uploading a large batch of vacation videos without labeling them individually, the tags will more likely be general descriptors (``Barcelona 2014'') rather than descriptions of specific activities in the videos. (Though for events like Ev101 {\it Birthday Party} that tend to cover a longer timespan with less unified activities, batch-labeled videos may be more likely.) It is not known whether batch-labeled videos have any particular characteristics in common that a search-collected video corpus would be skewed away from.

As well as the language of the video being familiar to (at least one of) the corpus-builders, there are other characteristics of videos that might make it easier to decide whether a target event is occurring, such as explicit internal descriptions of what's happening, like narration or titles. Length may also have an effect on interpretability; the average length for the YLI-MED positive videos is \~{}45 seconds, while the average length is \~{}40 seconds for YFCC100M as a whole.\fn{This figure is interesting, but not conclusive, given that videos of less than 3 seconds were explicitly excluded from the YLI-MED positives even in those rare cases where they might have been interpretable to the annotators.}

While quantitative testing of these speculations would require an entire dataset annotated for the relevant features, some preliminary insights can be gained from the data described in \S\ref{sec:psych}. For example, post-production text (like titles and subtitles) did not significantly affect YLI-MED annotators' stated confidence in their judgments nor agreement between annotators, but whether or not the language (spoken or text) was English affected both. Confidence and annotator agreement do not necessarily predict skew in corpus composition, but may provide an indirect indicator.

Finally, there are some biases that would not be helped even by individually reviewing every video on the Internet for possible corpus inclusion. The most obvious is against videos with a low-quality signal in one or both data streams, making it difficult to determine whether an event is in fact occurring (especially if there is no supporting metadata). However, while it is generally helpful for classification algorithms to be able to deal with somewhat low-quality videos even for consumer applications, if the quality is so poor that the humans constructing the corpus can't identify the event, it is unlikely to be a useful retrieval result in any case. 

\subsection{Intentional Compositional Differences from HAVIC}
\label{sec:havicdiffs}

Some differences between the characteristics of YLI-MED and those of HAVIC/TRECVID MED were intentional desiderata, not just chance consequences of procedure or base dataset choice. 

For example, HAVIC does not include animated videos/CGI (nor screencasts), which are always constructed (never natural) and present a different set of challenges for visual analysis than live recordings. %
However, HAVIC {\it does} include videos that have no audio track at all, which are, of course, simply unanalyzable in the audio stream. For YLI-MED, we required that positive examples (and near misses/related videos) have data in both the audio and visual streams---i.e., that they be \textit{multi}media.\fn{It is possible that this choice also introduced unintended differences, in that videos without sound may be biased in some other direction---towards lower-quality cameraphones, perhaps---%
though the effect is likely negligible at the scale of the whole corpus.}  However, we did not impose any requirements about the {\it type} of data in either stream, as long as one or the other provided evidence for the event---i.e., as long as it would likely be considered a good retrieval result by a user. (Both HAVIC and YLI-MED include videos where the visual track is live video (or a photo montage) but the audio consists only of a music track. In such cases, the audio data represents a separate event (i.e., a music performance) from the original live recording we are targeting, but it is nonetheless data.)

Similarly, our choice to include videos with non-human protagonists was intentional. Our goal is to improve video retrieval for the general public, who might well consider a video depicting robots or anthropomorphized animals celebrating a birthday to be a perfectly good example of a birthday-party video. It is not clear what consequences this might have for classification, though one can imagine some possibilities: for example, the colors and shadows on a robot's face reflecting a birthday-candle flame might not be detected in the same way that a human's would. However, it should be noted that the events targeted for YLI-MED Version 1 most often have human protagonists, and thus the vast majority of the videos do feature humans in the leading roles.

Finally, while it is not a difference in likely corpus composition, it is worth mentioning here that the additional annotations (beyond the event ID) chosen for HAVIC and YLI-MED differ according to feasibility and to the interests of the likely users of the corpora. Most importantly, we include in YLI-MED explicit annotations for each video about degree of annotator agreement and about the average annotator confidence in their judgments; we invite users to experiment with how using this information in various ways affects classifier performance and human perception of the usefulness of retrieval results.

\subsection{In the End, Is YLI-MED Comparable to Similar Corpora?}
\label{sec:compclassifiers}

To get some initial hints about how choice of base dataset and differences in collection procedures might actually affect research using the two corpora, we performed some rough experiments comparing YLI-MED Version 1 with the parallel sets of HAVIC videos used in TRECVID MED 2011 and later evaluations. Restrictions on the use of HAVIC data prevent us from making a direct comparison, i.e., training am event-detection system on one set and testing on the other. 
However, we can at least compare accuracy results for training and testing a system on YLI-MED Version 1.2 with parallel results for training and testing the same system on TRECVID MED data when we participated in the evaluations that included those events. 

The system we chose for the comparison is based on an i-vector approach initially developed by \citeauthor{Dehak2009} (\citeyear{Dehak2009}), with an improvement by \citeauthor{Burget2011} (\citeyear{Burget2011}). It involves training a matrix $T$ to model the total variability of a set of statistics for each video's audio track, primarily the first-order Baum-Welch statistics of the low-level acoustic feature frames (namely, Mel frequency cepstral coefficients). The i-vectors for each event class are averaged, so that each class is represented by one i-vector, then a generative Probabilistic Linear Discriminant Analysis (pLDA) (\citealt{plda}) log-likelihood ratio is used to obtain a similarity score between each test audio and each training event class. 
Full details about the i-vector system and some of our research using it with the TRECVID MED data may be found in \citealt{ElizaldeISM13}.

In addition to the work described in \citealt{ElizaldeISM13}, we used the i-vector approach to provide audio analysis for SRI-Sarnoff's AURORA multimedia event detection and retrieval system when it competed in TRECVID in 2012 and 2013 (\citealt{aurora2012,aurora2013}).  Table~\ref{tab:tvcomp} compares the event-detection performance of the i-vector system trained and tested on the TRECVID MED 2011 and 2013 data vs.\ the same system trained and tested on YLI-MED.\fn{\citealt{aurora2012} and \citealt{aurora2013} do not report individual results from each of the subcomponents of the AURORA system. The numbers shown in Table~\ref{tab:tvcomp} represent the accuracy for the i-vector system alone, when it was used in the experiments described there. The accuracy data for Attempting a Board Trick, Feeding an Animal, Landing a Fish, Wedding Ceremony, and Working on a Woodworking Project are from 2012 experiments that used the TRECVID MED 2011 DEV-T dataset, while the accuracy data for Birthday Party, Flash Mob Gathering, Getting a Vehicle Unstuck, Parade, and Grooming an Animal are from 2013 experiments that used the TRECVID MED 2013 MEDTest dataset.} In each case, there were two experimental runs, one with negatives and one without negatives.

\begin{table*}[htp]
\centering
\includegraphics[width=1\columnwidth]{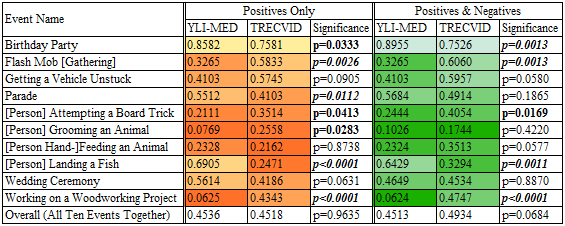}
\caption{Accuracy scores for an i-vector event detection system trained and tested on YLI-MED Version 1.2 vs.\ the TRECVID MED 2011 and 2013 datasets, and statistical significance of the differences between scores for each event. (Colors indicate scaled accuracy rankings within each experimental condition. Statistically significant differences are shown in boldface italics, with marginally significant differences in boldface alone.)}
\label{tab:tvcomp}
\end{table*}

It is important to note at the outset that there were many factors at play, and it is difficult to say how much the results shown in Table~\ref{tab:tvcomp} tell us about the audio characteristics of the two datasets as opposed to how much they tell us about other aspects of the experimental conditions or about the robustness and generalizability of the i-vector approach. In addition to the fact that the event sets are (largely) composed of different videos (determined in part by the factors mentioned in mentioned in \S\ref{sec:biasesdisc} and \S\ref{sec:havicdiffs}), they also differ in the total numbers of training and test files for each event. Other factors were more extrinsic to the event set makeup, such as the fact that the experiments with the TRECVID MED data involved 15 or 20 events rather than 10, so the range of possible (mis)classifications was different.\fn{This discrepancy in the total number of possible categories makes it more convenient to present results here in terms of accuracy scores, which are easier to compare across such different cases than mean average precision (MAP) or other metrics more often used in the TRECVID community.} In addition, the ratio of negatives to positives is much higher for YLI-MED Version 1, at around 27:1, as opposed to 7:1 for TRECVID MED.

Given these factors, we cannot use these data to make definitive statements about the similarity of the two corpora, even within the audio domain. However, we can at least sketch a few observations. First, the combined accuracy scores for the ten events were quite similar between the TRECVID and the YLI-MED runs within each condition (positives only vs.\  positives and negatives).\fn{Note that the overall accuracy scores given for TRECVID combine the scores for only these ten events, not the overall scores for each whole run, which included other events.
} The small difference between the scores for the positive-only runs was not at all statistically significant---though for the runs with both positives and negatives (which introduced even more differences in the overall experimental conditions), the difference was on the threshold of significance (Fisher's exact test, two-tailed). 

However, looking at the individual event accuracies that make up that overall score, we see a rather different picture. Looking first at the gross accuracy for each event, we find that detectors trained on the two corpora performed rather differently (at least when tested on videos from those same corpora). Those differences meet the threshold for statistical significance for most of the individual events for the positives-only runs, and for about half of the events for the runs with both positives and negatives. In addition, even where the differences do not meet the threshold of significance for assuming the scientific validity of the apparent effect, the p-values are often still rather low, i.e., the likelihood is still fairly high that the effect is {\it not} just random chance. (In fact, none of the p-values for individual events exceed 0.9.)

There is a greater range in accuracy across events for detectors trained and tested on YLI-MED data than TRECVID data. This wider scatter necessarily introduces wider differences between the gross accuracies that might obscure similarities in the overall patterns of which events the system performs better and worse on when trained and tested on the two datasets. We therefore also examined the {\it rankings} of the accuracy scores for the different events under each experimental condition; these (scaled) rankings are represented by shading in Table~\ref{tab:tvcomp}. 

For example, the i-vector system did best at detecting birthday parties on both corpora and under all conditions, while animal grooming was near the bottom for both---even though the gross accuracy scores for {\it Birthday Party} showed a statistically significantly difference. However, on the whole, the matches and near-matches in event rankings for the positives-only runs do not rise above what would be predicted for two randomly ordered lists.
For the runs with both positive and negative videos, the similarities in ranking patterns are more robust (rising above chance), with three exact rank matches ({\it Birthday Party}, {\it [Person] Attempting a Board Trick}, and {\it [Person Hand-]Feeding an Animal}), as well as an additional two events for which the accuracy rankings are off by only one step and one  additional event for which the ranking is off by two steps.
Looking at broad groupings rather than exact rankings, we find that the accuracy results for the positives-only runs fell into the same trentile for only three of the events, which would be predicted for random chance. In the runs with both positives and negatives, accuracies fell into the same trentile for six events, a bare majority. 

Taking all these patterns together, the i-vector system appears to perform somewhat similarly in terms of which events it is better and worse at detecting in the two datasets when negative examples are included. When negatives are not included, there do not seem to be even such weak inter-event patterns. 

Again, without being able to perform a more direct comparison of detectors trained on the two datasets, it is difficult to determine the sources of difference.\fn{We cursorily examined a few obvious potential factors, such as the numbers of videos in the test sets from the two corpora and the difficulty of finding videos for those events for YLI-MED, but found no obvious correlation with similarity in gross accuracy nor similarity in accuracy rankings.} The discussion thus far assumes that the training and test videos within a given corpus have close to the same distribution of audio characteristics for each event
(though see \S\ref{sec:ttcomp}). But it is possible that disparities in performance for the same system learning to detect the same event on the two different corpora may simply be an effect of there being more difference between the training and test sets for one corpus than the other. 

These i-vector experiments are, of course, only one means of getting at this question. At the least, it is quite possible that other audio-analysis approaches would be more or less robust to the differences between corpora, and detection systems using the visual-stream data from the two corpora might produce quite different results. However, we can at least tentatively say that the two corpora, though both relying on UGC and constructed with relatively similar desiderata, are disparate in ways that affect detection performance. 

The general assumption in the multimedia-retrieval community has been that, as corpora get bigger, the models trained on them will get more generalizable. However, according to the results described here, we cannot assume that event models trained on either YLI-MED or TRECVID would be generalizable to other UGC video corpora. Thus, although we may be making progress on generalizability, even corpora as large as YLI-MED and TRECVID MED are not yet sufficient for this task. 

We are currently working on a number of additional experiments to quantify the generalizability of classifiers trained on the YLI-MED, in tandem with testing the scalability of new classification and detection approaches. The YLI corpus includes multimedia-analysis tools being created at ICSI and University of California--Berkeley, including audioCaffe (\citealt{audiocaffe2015}), a set of audio content-analysis tools based on the deep neural net (DNN) image-analysis framework Caffe (\citealt{jia2014caffe}). We are training DNNs on the YLI-MED corpus, using the MFCC audio features (Mel frequency cepstral coefficients), and comparing them with the i-vector results to explore trade-offs between accuracy and efficiency  (\citealt{audiocaffe2015})---and meanwhile evaluating the comparability of YLI-MED and the TRECVID MED data.

\section{The Bigger Picture for YLI-MED}
\label{sec:future}

Here, we discuss how the YLI-MED annotations fit in with other current and planned efforts to make the YLI corpus a maximally useful resource for a broad sector of multimedia researchers, including the ongoing extraction of audio and visual features (\S\ref{sec:features}) and future annotation efforts (\S\ref{sec:nextsteps}).

\subsection{Feature Computation}
\label{sec:features}

The goal of the YLI corpus is to provide researchers with a multitude of pre-computed data to work with for each of the media files in the Yahoo Flickr Creative Commons 100 Million, to allow research teams to focus on conducting experiments rather than on computing statistics and features. Researchers at Lawrence Livermore National Laboratory and ICSI have been computing visual, audio, and motion features commonly used in multimedia analysis, using LLNL's Cray Catalyst, and making them available to the public. The visual features span the gamut of global (e.g., Gist), local (e.g., SIFT), and texture (e.g., Gabor) descriptors; the audio features include power spectrum (e.g., MFCC) and frequency (e.g., Kaldi) descriptors; and the motion features will begin with dense trajectories (\citealt{YLIfeatures}).

Audio features and LIRE image features for all of the videos in YFCC100M have already been released.\fn{\url{http://www.yli-corpus.org/computed-features}} Many more visual and motion features are scheduled to be released over the coming months; the full list includes: 

\bi
\item Audio features: \bi
\item MFCC20s (Mel frequency cepstral coefficients, 19 channels plus energy)
\item Kaldi pitch
\item SAcC pitch (subband autocorrelation classification)\ei
\item Image features on video keyframes: \bi
\item Gist features
\item LIRE features: auto color correlogram, basic features, CEDD (color and edge directivity descriptor), color layout, edge histogram, fuzzy color and texture histogram (FCTH), fuzzy opponent histogram, Gabor features, joint histogram, joint opponent histogram, scalable color, simple color histogram (RGB), Tamura texture
\item AlexNet features
\item SIFT features\ei
\item Motion features: \bi
\item Dense trajectory features \ei
\ei

\noindent A subset of the YLI corpus data, including the computed features, was already used in the MediaEval 2014 Placing Task, and additional YLI data will be added in 2015 and 2016 (\citealt{MP14}).

We are also releasing convenient bundles of features just for YLI-MED, ahead of the releases for the full YLI corpus. A first bundle of features, including Kaldi pitch, SAcC pitch, and MFCCs (audio) and AlexNet and LIRE (image) for (most of) the positive-example videos in YLI-MED, was released in November 2014 and is available at \url{http://www.yli-corpus.org/the-yli-med-corpus}. A full bundle for those features, covering all of the videos in YLI-MED, will be available in late spring 2015, with additional features continually being added.

Details of the YLI feature-computation procedures and technical basis can be found in \citealt{YLIfeatures} and \citealt{MP14}. While the features for YLI-MED were computed separately, the organization and rationale mirror that of the full YLI feature corpus.

\subsection{Next Steps for YLI-MED}
\label{sec:nextsteps}

We are currently moving towards launching a full-fledged ``Multimedia Genome Project'' (MMGP), in collaboration with Lawrence Livermore National Laboratory. In the MMGP, we hope to provide reliable human-generated labels for all of the \~{}800,000 videos in YFCC100M, using a combination of crowdsourcing and expert annotation. The MMGP will include a heavy emphasis on multimedia event annotation, including loose annotations for the whole corpus as well as defined sets of training and test data for specific well-defined events like those described here. In particular, we hope to add twenty to fifty new events to the YLI-MED, including new events never seen in TRECVID.

We anticipate that having such a large-scale, open-source annotated corpus of user-generated videos will truly be a game-changer for multimedia research. The MMGP will provide a basis for collaboration and exploration that pushes the boundaries of multimedia analysis and retrieval; for example, it has the potential to enrich our understanding of the relationship between more high-level characteristics like multimedia events and more concrete, localizable characteristics like audio concepts. We hope such a project will lead to new methods of inquiry that would not otherwise be possible, encourage the development of new data analysis tools and algorithms, and support students and new researchers who might not otherwise have access to large, high-quality video corpora.

We have already received a great deal of interest in the projected MMGP effort from leaders in event detection as well as other multimedia subfields. In the long term, it has the potential to become the basis for major collaborative research and evaluation efforts. The idea is also of interest to commercial multimedia research and development, and we are already working on developing collaborations that could support better access to the YLI data, including YLI-MED.

In the short term, the next step is to develop and release Version 2 of YLI-MED, with at least five new events. In the first round of annotations, we chose TRECVID MED events in order to 
provide a basis for comparison for researchers in that community. 
In the second round, we will branch out and explore an area that TRECVID specifically does not, focusing on events that prototypically have non-human protagonists. In this second round, we will likely choose a more closely related set of events within a single conceptual domain, offering a different set of challenges for classification than more distantly related events. 
We will also include new non-event annotations for additional important video characteristics, such as narration, content text, and use of static images.

\section{Acknowledgments}
\label{sec:ack}

This annotation effort is vitally connected to the work of the researchers at ICSI and Lawrence Livermore National Laboratory who are computing the audio, visual, and motion features for the videos in the YLI-MED corpus, including Jaeyoung Choi (ICSI) and Carmen Carrano, Karl Ni, Roger Pearce, and Doug Poland (LLNL), in addition to some of the authors. 

YLI-MED would not have been possible without the efforts of the researchers at Yahoo Labs who produced the YFCC100M dataset in the first place, including Bart Thomee, Li-Jia Li, Nikhil Rasiwasia, and David Ayman Shamma, along with the Yahoo Webscope and Legal teams.

The YLI-MED annotation project described here was funded by a grant from Cisco Systems, Inc., for Event Detection for Improved Speaker Diarization and Meeting Analysis. In addition to Cisco, ICSI's work on the YLI corpus (generally) is funded by Lawrence Livermore National Laboratory as part of a collaborative Laboratory Directed Research and Development project under the auspices of the U.S.\ Dept.\ of Energy (contract DE-AC52-07NA27344) and by the National Science Foundation as part of SMASH: Scalable Multimedia content AnalysiS in a High-level language (grant IIS-1251276). Any opinions, findings, and conclusions or recommendations are those of the authors and do not necessarily reflect the views of Cisco, LLNL, nor the NSF.

Questions and requests for further information may be directed to  \ylicontact.

\bibliographystyle{named}
\bibliography{main}

\appendix

\section{Event Definitions for the YLI-MED Corpus}
\label{sec:defsapp}

\subsection*{Introduction for the Corpus-Builders\fn{This introduction and definitions were originally intended as instructions for the video collectors and verifiers who built the YLI-MED corpus. They are framed for that purpose, and assume a certain amount of background knowledge. Copies of the full instructions provided to the corpus-builders, as well as definitions for the five events that were not included in the final corpus, may be obtained by emailing \ylicontact.}}

The event definitions/schemas and decision rubrics in this document are intended to help you identify \textit{positive instances} of each type of event. Where a rubric says something ``does not qualify'' or ``should not be included'', that means it should not be included as a positive instance; however, it might well still qualify as a `near miss' or a `related video' for that category.
We assume it will be relatively easy for you to make decisions about the most prototypical instances of a given event type, and more difficult to make decisions about the less prototypical instances (at the boundaries of the category).\fn{Not required reading, but might be fun: \url{https://en.wikipedia.org/wiki/Prototype_Theory}.} The bulk of the decision rubrics are therefore devoted to what to do with the boundary cases, rather than the easy cases that will probably be most of what you find.

\textit{What's an event?} For the purposes of this task, an event ``is a complex activity occurring at a specific place and time; ... consists of a number of ... actions, processes, and activities that are loosely or tightly organized and that have significant temporal and semantic relationships to the overarching activity; and is directly observable [via at least one sensory modality]'' (modified from NIST’s definition of an ``event'' for the TRECVID MED evaluations\fn{\url{http://www.nist.gov/itl/iad/mig/med11.cfm}}).

\textit{What’s a person?} A ``person'' for the purposes of these definitions includes not only humans but also other agentive beings,\fn{For these purposes, an agentive being is an animate being or entity---or a being or entity that can be construed as animate---that is understood as having the capacity to choose its actions intentionally.} if they are obviously intended to be anthropomorphic in some way. In live video, ``people'' may include animals or robots, if the creator of the video and/or the participants are obviously anthropomorphizing them (this may include animals trained to mimic human behaviors). In an animated video, ``people'' may include almost anything, if they have human-like behavior and speech.

\subsection*{\textit{Event Name:} Birthday Party (Ev101)}

\textbf{\textit{Definition:}} Multiple people gather for a social event whose purpose is to celebrate someone’s birthday, by engaging in some special activities that are intended to be enjoyable.

\noindent \textbf{\textit{Decision Rubrics/Clarifications:}}
\bi
\item A \textit{party} is a specific type of social gathering where the goal is for everyone present to enjoy themselves more than they usually would in day-to-day life by doing something generally viewed as special; this generally involves some activities that participants wouldn’t engage in every day, or at least some special spin on regular activities.
\item One person does not a party make. (However, one person and one animal might.)
\item The prototypical American birthday-party elements like birthday cake with candles and people singing ``Happy Birthday'' are not required, as long as the general definition of a party is fulfilled and it is clear that a birthday is what is being celebrated.
\item It is not necessary for the person whose birthday is being celebrated to be present, or even alive. The birthday does not have to be that of a human; it can be an animal’s birthday or even an anniversary of the founding/beginning of some organization/entity, as long as it is explicitly being treated as a ``birthday'' by the participants in the party. 
\ei

\subsection*{\textit{Event Name:} Flash Mob\fn{ Video collectors and verifiers were initially given the name \textit{Flash Mob Gathering} for this event.} (Ev102)}

\textbf{\textit{Definition:}} Multiple people gather in a public place and perform some type of prearranged coordinated action that is particularly unusual to be performing in that situation and/or place, generally for the purpose of surprising others present. 

\noindent \textbf{\textit{Decision Rubrics/Clarifications:}}
\bi
\item The prototypical purpose of a flash mob is to surprise passersby (usually pleasantly) and give both participants and nonparticipants a break from their routine, perhaps to shake things up a bit or poke fun at social conventions. The activity often involves artistic performance (such as dance or music) and/or some actions that would normally be considered socially inappropriate for the setting, or at least a mismatch for the setting. However, flash mobs may also have other purposes like making a political statement or may be part of a publicity campaign; videos depicting such gatherings should be included as long as they are prototypical flash mobs in all other dimensions.
\item A flash mob may begin in a couple of different ways: the participants suddenly arrive and gather as a group in the designated place, then begin the prearranged action, or the participants gradually filter into the designated place and mingle with the crowd, then suddenly begin the prearranged coordinated action from their positions interspersed among nonparticipants. Flash mobs are usually relatively short (on the scale of a few minutes) and usually end with the participants dispersing quickly, but these are not strict requirements.
\item A flash mob may be organized in a couple of different ways: The participants may all know each other and may even have practiced the planned action together before, or they may not know each other, but be summoned to perform the action by some organization or person who has their information on a contact list. The participants may know the planned time, location, and/or activity for the flash mob well ahead of time, or they may receive a message (text or otherwise) with the information just when it is time for the action to start, telling them to go immediately if they want to participate.
\item Special note: The concept \textit{flash mob} is a ``contested category'', in that reasonable people frequently disagree about where the boundaries are and what should count as a ``real'' flash mob. For the purposes of this task, you can allow the definition to stretch pretty broadly.
\ei

\subsection*{\textit{Event Name:} Getting a Vehicle Unstuck (Ev103)}

\textbf{\textit{Definition:}} A person or persons perform(s) some actions with the goal of getting some type of vehicle (motorized or unmotorized) out of a situation where it is stuck, i.e., either cannot move at all or is severely limited in its movements, into a situation where it can move normally.

\noindent \textbf{\textit{Decision Rubrics/Clarifications:}}
\bi
\item The attempt does not have to be successful; however, someone must have an obvious intention to get the vehicle into a position or state where it can be moved normally.
\item A \textit{vehicle} is any mechanical device (motorized or unmotorized) that can be used to transport people or objects by means of moving said device. Vehicles that move people or objects any distance count, from within-building vehicles (like shopping carts) to intercontinental vehicles (like airplanes). 
\item The vehicle’s movement may be impeded either because some part of it is embedded in some substance that is preventing it from operating normally (usually something that prevents the wheels from turning or from getting purchase), or because it is situated at an angle with respect to surrounding objects that makes it impossible for it to exit the space in a normal trajectory, or any other impeding situation with similar results. A vehicle does not count as \textit{stuck} if the impediment is necessary to normal parking or storage of the vehicle (e.g., getting a bike off a bike hook does not count).
\item The activities may involve moving the vehicle itself or may involve moving other obstacles or impediments.
\ei

\subsection*{\textit{Event Name:} Parade (Ev104)}

\textbf{\textit{Definition:}} A group of people moves in a procession formation through a public space, under their own power or in vehicles, while simultaneously engaging in special activities or exhibiting special behaviors to signal they are celebrating something.

\noindent \textbf{\textit{Decision Rubrics/Clarifications:}}
\bi
\item A parade is organized---or may spontaneously arise---to celebrate something (like a holiday or an event). Public processions organized for other purposes (such as protest marches, religious processions, or funeral processions) generally do not qualify. However, processions with explicit religious, funereal, or political-oppositional purposes may count if they resemble the prototypical parade in style and celebratory mood.
\item Standard celebration-signaling elements like music and colorful signs/banners/flags are not required as long as there are some type of obvious markers that the procession is a special celebratory event.
\item Videos showing the participants forming up for the procession, i.e., arranging themselves into the planned order/configuration, can be included as part of the parade event, even if the video does not depict the procession itself, as long as the intention is obvious and clear. (Other preparatory steps do not qualify, if the procession itself is not depicted.) Videos depicting parade participants stopping to interact with spectators should be included, but videos that depict only spectators should not.
\ei

\subsection*{\textit{Event Name:} Person Attempting a Board Trick\fn{Video collectors and verifiers were initially given the name \textit{Attempting a Board Trick} for this event.} (Ev105)}

\textbf{\textit{Definition:}} A person or persons attempt(s) to perform a maneuver requiring special skill or practice, on some type of board-like object designed to allow a human to move swiftly across a surface, usually under their own power.

\noindent \textbf{\textit{Decision Rubrics/Clarifications:}}
\bi
\item The attempt does not have to be successful, nor does it have to be unsuccessful (either way counts). However, there must be an obvious intention to perform the trick on the part of the performer.
\item The attempted maneuver must be a special action requiring skill and/or practice to precisely control the motion of one’s body and the board to achieve the intended effect; the sort of thing one might do to surprise, amuse, or impress an audience (even if there is no audience in the video).
\item The prototypical board trick involves a skateboard, snowboard, or surfboard. Other flat boards may qualify, if they are specially designed to allow a person (usually one person per board) to move quickly across a surface of some kind, to which the face of the board would usually be parallel, where the person could maintain a relatively static body position relative to the board while it moved; or if they are obviously being used as stand-ins for specially designed boards. Boards with supplemental (non-human) power sources may qualify if they are very closely related to the prototypical types of boards (for example, electric skateboards, or a fantasy equivalent in an animation). Toy boards that are obviously intended to imitate the prototypical types of boards may count if the attempted trick resembles a prototypical board trick in all other dimensions.
\ei

\subsection*{\textit{Event Name:} Person Grooming an Animal\fn{Video collectors and verifiers were initially given the name \textit{Grooming an Animal} for this event.} (Ev106)}

\textbf{\textit{Definition:}} A person or person(s) performs some type of external body care on a (nonhuman) animal to make it cleaner and/or to improve its appearance.

\noindent \textbf{\textit{Decision Rubrics/Clarifications:}}
\bi
\item Grooming prototypically involves caring for an animal’s skin, coat/fur, feathers, scales, etc., but any type of (external) body care qualifies if its purpose is to make the animal cleaner and/or neater or more attractive in appearance (according to some standard set by the groomer).
\item Animal care focused on the animal’s skin, fur, etc. that is not for the purpose of making it cleaner or more attractive (for example, petting or applying topical medicines) does not count as grooming. 
\item The grooming must be performed by (a) human(s) or anthropomorphized agent(s).  
\item In animations, the animals may be invented beasts that do not match any species in the kingdom Animalia in reality, as long as they have the general characteristics of nonhuman animals and are obviously intended by the creators of the video to fit that general category/fulfill a role analogous to the role of animals on (real) Earth.
\ei

\subsection*{\textit{Event Name:} Person Hand-Feeding an Animal\fn{Video collectors and verifiers were initially given the name \textit{Feeding an Animal} for this event.} (Ev107)}

\textbf{\textit{Definition:}} A person or persons intentionally provide(s) food and/or water for some (nonhuman) animal(s) to eat, by holding the food or water or a container with food or water in it while the animal eats it.\fn{It should be noted that this definition implies that, say, tossing food to fish or birds would not count, even if the person doing the feeding remains present and engaged and continues providing food as the animal is eating. This omission was not corrected explicitly, but some of the corpus-builders did count such videos as positive instances.} 

\noindent \textbf{\textit{Decision Rubrics/Clarifications:}}
\bi
\item The feeding must involve direct interaction between the person doing the feeding and the animal(s) being fed; putting food somewhere where the animal will find it does not count.
\item Any type of solid or liquid food can be provided (including ``treats''), whether or not it is actually appropriate for the animal's diet. 
\item The attempt at feeding does not have to be successful, in that the animal(s) do(es) not have to actually accept the food in the course of the video, as long as it is clear that the person doing the feeding intends that the food should be eaten by the animal(s).
\item In animations, the animals may be invented beasts that do not match any species in the kingdom Animalia in reality, as long as they have the general characteristics of nonhuman animals and are obviously intended by the creators of the video to fit that general category/fulfill a role analogous to the role of animals on (real) Earth.
\ei

\subsection*{\textit{Event Name:} Person Landing a Fish\fn{Video collectors and verifiers were initially given the name \textit{Landing a Fish} for this event.} (Ev108)}

\textbf{\textit{Definition:}} A person or persons who has/have successfully caught a fish with some implement use(s) the implement to bring the fish in from the water to where they are situated.

\noindent \textbf{\textit{Decision Rubrics/Clarifications:}}
\bi
\item \textit{Landing} a fish involves not only catching it with the implement (getting indirect control of it) but also getting it out of the water and into the immediate proximity/direct control of the fisher (i.e., successfully reeling it in, for rod fishing). Videos that depict only the fish getting caught on/in the fishing implement should not be included; they must depict at least an attempt to bring the fish in from the water. 
\item The fish does not necessarily have to be brought to dry land; it can be ``landed'' on a boat or a dock (or even in the hands of a person standing in the water), as long as it is in a place where it would not end up back in the water if it were released from the implement that was used to catch it.
\item The fish must be caught from a body of water (aquariums don’t count).
\item The fisher must be a human or anthropomorphized agent who is actively involved in operating the implement to effect the catch and landing (hauling in traps does not count).
\item The prototypical implement is a fishing rod and line, but nets and rodless lines may be used as long as they can catch only a few fish at a time and the overall activity is focused on the challenge of catching each individual fish. 
\ei

\subsection*{\textit{Event Name:} Wedding Ceremony (Ev109)}

\textbf{\textit{Definition:}} Multiple people participate in a ceremony that includes elements that result in two or more of the participants being considered married according to some standard (legal, societal, religious, and/or personal).

\noindent \textbf{\textit{Decision Rubrics/Clarifications:}}
\bi
\item For these purposes, a \textit{ceremony} is an event in which a person or persons perform(s) a formal sequence of prescribed actions and/or words that is viewed by that person or persons as appropriate to and definitional of that event, usually on the basis of the conventions of a larger community or society.
\item The prototypical Western/Judeo-Christian wedding elements like rings, formal vows, or even an officiant with legal or religious authority are not required, as long as it is clear that at least some of the participants in the ceremony believe and intend that their performance of the prescribed words and/or actions will result in a marriage between two or more humans or anthropomorphized agents (who may or may not themselves be present at the ceremony).  
\ei

\subsection*{\textit{Event Name:} Working on a Woodworking Project (Ev110)}

\textbf{\textit{Definition:}} A person or persons perform(s) steps in the process of making something out of wood using tools, as part of a specific project.

\noindent \textbf{\textit{Decision Rubrics/Clarifications:}}
\bi
\item A \textit{project} is a single unified undertaking with a specific planned result or end condition. The prototypical woodworking project is undertaken casually, as a hobby or to fulfill a household need, but a woodworking activity could still qualify as a project if it were undertaken for pay in some circumstances: if the woodworker has creative and economic control of the production, or if their job usually does not involve woodworking. However, woodworking as part of a daily job with no creative control, as in industrial production or most construction, does not qualify as a \textit{project}.
\item \textit{Woodworking} usually refers to making furniture or smaller decorative or functional objects, or to cabinetry, but videos may be included that depict someone making wood elements for larger-scale projects such as building construction, if the process qualifies as a prototypical woodworking project in all other ways. Videos of someone repairing something made of wood may be included if it requires specialized knowledge or tools.
\item The central aspect of \textit{woodworking} is using tools to cut and shape wood, but videos should also be included that depict other steps necessary for producing an item made out of wood, including measuring wood, staining or decorating wood, and assembling pieces of wood and other materials together to make a finished product. (However, planning, design, and acquiring materials do not count as part of working on the project.)
\ei

\end{document}